\documentclass[bibyear]{aa}  

\usepackage{natbib}
\usepackage[breaklinks=true]{hyperref}
\bibpunct{(}{)}{;}{a}{}{,} % to follow the A&A style
\usepackage{graphicx}
\usepackage{txfonts}
\usepackage[normalem]{ulem}
\usepackage{amsmath}
\usepackage{multicol}

\usepackage{color}
\definecolor{myorange}{rgb}{0.8, 0.3, 0.0}
\definecolor{myblue}{rgb}{0.144, 0.55, 0.82}
\definecolor{lbgreen}{rgb}{0.0, 0.42, 0.24}
\usepackage{soul}

\newcommand{\yrs}{\,\mathrm{yrs}}

\newcommand{\Modot}{\,\mathrm{M_\odot}}

\newcommand{\lgbh}{\emph{L-Galaxies}\texttt{BH}}
\newcommand{\lgfamily}{\emph{L-Galaxies}}
\newcommand{\phaseflow}{\textsc{PhaseFlow}}
\begin{document} 

%SINGLE TITLES
%\title{Demographics of Tidal Disruption Events and Constraints on the Compact Object Occupation of Galactic Nuclei}

%\title{TDEs constraining the compact object occupation of galaxies}

%TITLE + SUBTITLE
% heavy one 
%\title{Tidal Disruption Events from simulated Massive Black Holes; modeled, not assigned}
%\title{A cosmological volume full of Tidal Disruption Events}
%\title{Tidal Disruption Events within a Cosmological Box of Massive Black Holes}
%\title{Tidal Disruption Events within a Small Universe of Massive Black Holes}
\title{Demographics of Tidal Disruption Events with \texttt{L-Galaxies}}
\subtitle{I. Volumetric TDE rates and the abundance of Nuclear Star Clusters}

%\subtitle{I. On the Compact Object Occupation of Galactic Nuclei and late-time black hole growth through star-accretion}

\author{\href{https://www.orcid.org/0000-0002-8889-2167}{M.~Polkas}\inst{1,2}\thanks{markos.polkas@ dipc.org} \and 
\href{https://www.orcid.org/0000-0002-6381-2052}{S.~Bonoli}$^{1,3}$ \and \href{https://orcid.org/0000-0001-9458-821X}{E.~Bortolas}$^{4,5}$ \and \href{https://www.orcid.org/0000-0002-8889-2167}{D.~Izquierdo-Villalba}\inst{4,5}  \and \href{https://orcid.org/0000-0003-4961-1606}{A.~Sesana}$^{4,5,6}$ \and \href{https://orcid.org/0000-0002-9076-1094}{L.~Broggi}$^{4,5}$ \and \href{https://www.orcid.org/0000-0001-8040-4088}{N.~Hoyer}$^{1,7,8,9}$ \and \href{https://orcid.org/0000-0002-9074-4833}{D.~Spinoso}$^{10}$}
\institute{Donostia International Physics Center, Paseo Manuel de Lardizabal 4, E-20118 Donostia-San Sebasti{\'{a}}n, Spain\and
University of the Basque Country UPV/EHU, Department of Theoretical Physics, Bilbao, E-48080, Spain\and
IKERBASQUE, Basque Foundation for Science, E-48013, Bilbao, Spain\and
%
% Elisa, Alberto, Luca
Dipartimento di Fisica `G. Occhialini', Università degli Studi di Milano-Bicocca, Piazza della Scienza 3, I-20216 Milano, Italy \and
INFN, Sezione di Milano-Bicocca Piazza della Scienza 3, I-20126 Milano, Italy\and
%Alberto
INAF - Osservatorio Astronomico di Brera, via Brera 20, I-20121 Milano, Italy\and
%
% Nils
Max-Planck-Institut f{\"{u}}r Astronomie, K{\"{o}}nigstuhl 17, D-69117 Heidelberg, Germany\and % Nils
Institute of Astronomy, Pontificia Universidad Cat{\'{o}}lica de Chile, Avenida Vicu{\~{n}}a Mackena 4690, Santiago, Chile\and % Nils
Universit{\"{a}}t Heidelberg, Seminarstrasse 2, D-69117 Heidelberg, Germany\and 
%Daniele
Department of Astronomy, MongManWai Building, Tsinghua University, Beijing 100084, China
}

\date{Received 2023; accepted 2023}

     \abstract
   { Stars can be ripped apart by tidal forces in the vicinity of a massive black hole (MBH), causing luminous flares known as tidal disruption events (TDEs). These events could be contributing to the mass growth of intermediate-mass MBHs, and new samples from transient surveys can provide useful information on this unexplored growth channel. This work aims to study the demographics of TDEs by modeling the co-evolution of MBHs and their galactic environments in a cosmological framework. We use the semi-analytic galaxy formation model {\lgbh}, which follows the evolution of galaxies as well as of MBHs, including multiple scenarios for MBH seeds and growth, spin evolution, and binary MBH dynamics. Time-dependent TDE rates are associated with each MBH depending on the stellar environment, following the solutions to the 1-D Fokker Planck equation solved with {\phaseflow}. Our model produces volumetric rates that are in agreement with the latest optical and previous X-ray samples. This agreement requires a high occupation fraction of nuclear star clusters with MBHs since these star reservoirs host the majority of TDEs at all mass regimes. We predict that TDE rates are an increasing function of MBH mass up to ${\sim}\, 10^{5.5}\Modot$, beyond which the distribution flattens and eventually drops for ${>}\, 10^{7}\Modot$. In general, volumetric rates are predicted to be redshift-independent at $z\,{<}\,1$. We discuss how the spin distribution of MBHs around the 
 event horizon suppression can be constrained via TDE rates and what is the average contribution of TDEs to the MBH growth. 
 In our work, the majority of low-mass galaxies host nuclear star clusters that have their loss-cone depleted by $z\,=\,0$, explaining why TDEs are rare in these systems. This highlights that time-dependent TDE rates are essential for any model to be in good agreement with observations at all mass regimes.
   }

   \keywords{black holes --
                tidal-disruption-events -- nuclear-star-clusters --
                semi-analytic -- galaxy evolution
               }
   \maketitle
%
%-------------------------------------------------------------------

\section{Introduction}

A tidal disruption event (TDE) occurs when a star wanders too close to a massive\footnote{In the occurring field of TDEs from stellar mass black holes \citep{Kremer22,Kremer23,Vynatheya23,Xin24} the events are frequently referred to as micro-TDEs, so we preserve the term TDE for events from massive black holes.} black hole (MBH) so that the MBH gravitational pull overcomes the star's self-gravity. As a result, the star gets \textit{spagettified}, and a part of it settles into a disk-like configuration producing a distinct, multi-wavelength electromagnetic flare. TDEs can outshine their host galaxies, with luminosities of $10^{42}\,{-}\,10^{45}$erg s$^{-1}$ which decline over weeks to years timescales. Most TDEs can be identified by the characteristic post-peak decrease of their accretion rate, which drops for most of the events as $\,{\propto}\, t^{-5/3}$, as predicted by the standard fallback theory \citep{Rees88,Phinney89}. 

The first observations of TDE-like transients, initially in the X-ray \citep[e.g.][]{Bade96,Komossa99,Esquej08} and then in the optical/UV sky \citep{Gezari06,Gezari08,VanVelzen11}, sparked the interest on their overall rate per galaxy \citep{Magorrian99,Syer99,Wang04}. 
Such interest has grown even further at the present date, as the number of observed TDEs is growing faster than ever (we now have identified approximately 100 TDE candidates), mainly owing to the advent of wide-field transient optical surveys, such as Pan-STARRS1 \citep{Chornock14}, the Palomar Transient Factory  \citep[PTF, e.g.][]{Cenko12,Arcavi14},  the ongoing Zwicky Transient Facility \citep[ZTF, e.g.][]{Lin22,Hammerstein23,Yao23}, and ASAS-SN \citep[e.g.][]{Krolik16,Hinkle21,Liu23}. In the X-rays, eROSITA has already provided a sample of candidates \citep{Sazonov21} adding to previously compiled inventories from Swift, XMM-Newton, and Chandra \citep{Kawamuro16,Auchettl17, Goldtooth23}. Finally, individual dust-shrouded TDE candidates have been detected in mid-infrared \citep{Mattila18,Kool20,Wang22,Panagiotou23}, and a recent analysis of NEOWISE data has yielded the first sample at this wavelength \citep{Masterson24}.

% MAYBE WE SHOULD FIRST MENTION THE RATES AND COMPARE WITH OBSERVATIONS, THEN DISCUSS THE HOST GALAXIES FOLLOWING THIS
The collection of these growing samples has allowed us to assess the TDE rates. In particular, the works of \citet{vanVelzen18}, \citet{Lin22} and \citet{Yao23} used a relatively wide sample of observed TDEs to infer an overall volumetric rate of ${\sim}\, 10^{-7}\,{-}\, 10^{-6}$ Mpc$^{-3}$ yr$^{-1}$ $d \log_{10}\,L_{b}^{-1}$, with $L_b$ being the peak luminosity within the band of a given survey, corresponding to a number of TDEs per galaxy of the order of $10^{-5}\,{-}\,10^{-4}$ yr$^{-1}$. Although these results necessarily depend on the shape of the TDE luminosity function, which remains uncertain, the obtained rates are in decent agreement with (although somewhat lower than) the TDE rates predicted by theoretical and numerical studies \citep{Syer99,Magorrian99, Donley02, Wang04,Esquej08,Merritt09,Brockamp11,Stone16,Pfister21,Broggi22}. Nevertheless, both observational and theoretical estimates suffer from several limitations. On one side, the available sample of observed TDEs is still relatively small. On the other hand, for simplicity, theoretical models typically neglect important ingredients such as the time evolution of TDE rates under the evolution of the host galaxy. Most models are also affected by our poor knowledge of the MBH mass function at low masses \citep[${\lesssim}\,10^7 \Modot$, see][]{Shankar13,Greene20} and the occupation fraction of nuclear star clusters \citep[NSCs, see recent studies][]{Hoyer23,Ashok23,Hoyer24}.

%HOST-GALAXIES RELATION 
Still, currently available samples have allowed us to perform statistical studies on the host galaxies of observed TDEs (see \citealt{French20} for a review). In particular, ZTF observations have shed light on the fact that TDEs are over-represented in ultra-luminous Infrared Galaxies \citep[ULIRGs,][]{Tadhunter17,Reynolds22} as well as in the rarer post-starburst/green-valley galaxies \citep{Hammerstein21,Hammerstein23}, and in particular in the quiescent Balmer-strong (E+A) ones \citep{Arcavi14,French16,Law-Smith17,Graur18,Dodd21}. The stellar mass distribution of TDE-hosts is relatively flat compared to the stellar mass function and is concentrated in the dwarf-to-massive transition regime, ${\sim}\,10^{9}\,{-}\,10^{11} \Modot$ \citep{Wevers19}. Interestingly, the occupation fraction of NSCs peaks in the same mass range \citep[e.g.][]{Sanchez-Janssen19,Hoyer21}. 
That being said, MBHs and NSCs frequently have been related in many works \citep{Antonini15,Naiman15,Trani18,Askar22,Atallah23,Lee23,Tremmel23}, yet the exact nature of their connection remains unknown. What seems to be evident from theoretical studies \citep{Merritt09,Stone16,Pfister20}, is that the presence of an NSC, the densest stellar system possible in the universe \citep{Neumayer2020}, may enhance the rate of TDEs on the central MBH. Therefore, the majority of TDEs are expected to be related to intermediate-mass black holes (typically defined as $10^{2.5}\,{-}\,10^{6}\Modot$) in the center of NSCs. At the same time, the very existence of MBHs below $10^5\Modot$ is being challenged by the tentative and scarce observational evidence, especially towards the low end of the mass range (\citealp{Mezcua17}, but see recent robust evidence for an IMBH in Omega Centauri \citealp{Haeberle24}).

At this intermediate MBH-mass scale, the majority ($>$90\%) of black holes are believed to be inactive \citep{Greene20,Mezcua24}, making samples inferred from Active Galactic Nuclei (AGN) observations incomplete. Also, mass measurements through spectral information become troublesome at low masses \citep{Kormendy13}. In fact, beyond the local universe where accurate dynamical measurements of MBH mass can be made, TDE observations serve as the only \emph{direct} detection method of the dormant MBH population, as opposed to the indirect method of detecting relic AGN activity.

%TDEs as a growth channel  
Finally, TDEs offer a viable channel of black hole growth \citep{Hills75} that could in principle be dominant for MBHs that do not grow efficiently through gas accretion. After the first observations, the TDE growth channel was revisited \citep{Magorrian99}, with \citet{Milosavljevic06} proposing that low-mass MBHs ($\,< \, 2\,{\times} \,10^6 \Modot$) may acquire the majority of their mass through TDEs. Furthermore, \citet{Bar-Or17} used this channel to set a lower limit on the masses of MBHs that can exist in the local universe since all MBH seeds should grow either through gas or TDEs. The initial growth through TDEs has been also studied within zoom-in cosmological simulations \citep{Pfister21,Lee23} and, recently, high-resolution simulations achieved growing a black hole seed by a factor of $>20$ within 1{\,}Gyr \citep{Rizzuto23}. Yet, the argument of effective TDE growth has been questioned for low-mass galaxies, since NSCs which contribute mainly to the TDE rates are not dense enough to fuel black hole growth by runaway tidal capture of stars \citep{Miller12,Stone17}, a physical process that is more efficient at growing intermediate-mass MBH in globular clusters \citep{Portegies02,Arca-Sedda16} and hierarchically-merging star clusters \citep[see recent advancements in e.g.][]{Rantala24}. Nevertheless, a statistical study on the frequency of TDEs across a realistic population of stellar environments and the role of stellar accretion in the growth of MBHs has not been performed so far.\\

%In semi-analytic models of galaxy formation, galaxies have the simplest composition possible for fast calculations. On the other hand, even if the galaxies were included with the abundance of morphologies observed in the sky or cosmological simulation outputs where used, we would have needed an impossible amount of individual N-body simulations or direct Fokker-Planck approximation (orbital-average characterization is not longer valid) to assign to each set up a realistic time-dependent TDE rate. The simplicity of the current work aims to be efficient in capturing the dependence of the rates on time-evolving black holes and galactic environments on the fly without having to deal with all the specific configurations. 

In this paper, we set the foundation to address many of the aforementioned theoretical uncertainties by combining, for the first time, a semi-analytic model of galaxy/black hole evolution in a wide range of stellar environments with time-dependent TDE rates provided by a fast 1D-Fokker Planck solver. The paper is structured as follows; In Sect.~\ref{sec:model} we describe our method for coupling the physics of TDEs in a given local environment with a variety of stellar environments. In Sect.~\ref{sec:Results} we describe our most important findings and compare our TDE rate predictions with new constraints. In particular, we focus on the cosmological evolution of TDE rates and the contribution from active MBHs. In Sect.~\ref{sec:discussion}, we complement our analysis by addressing the impact of the parameter choice and discussing the implications for the occupation fraction of NSCs, the MBH spin model, and MBH growth. In Sect.~\ref{sec:lim_and_prosp}, we discuss some caveats and subjects that we aim to investigate explicitly in the future. Finally, we summarize the key aspects of our work in Sect.~ \ref{sec:conclusions}. Throughout the paper, we adopt a Lambda Cold Dark Matter $(\Lambda$CDM) cosmology with parameters $\Omega_{\rm m} \,{=}\,0.315$, $\Omega_{\rm \Lambda}\,{=}\,0.685$, $\Omega_{\rm b}\,{=}\,0.0493$, $\sigma_{8}\,{=}\,0.826$ and $\rm H_0\,{=}\,67.4\, \rm km\,s^{-1}\,Mpc^{-1}$ \citep{Planck2020}.

\section{Model Description}\label{sec:model}

In this work, we combine the semi-analytical model of galaxy formation {\lgfamily} with the 1D Fokker-Planck solver {\phaseflow} to estimate the time evolution of TDE rates and compare it with observations. In particular, we use a version of {\lgfamily}, dubbed as {\lgbh} throughout this work, developed to study a wide variety of physical processes that drive the evolution of the MBH population and its co-evolution with galaxies.  In this section, we first describe the {\lgfamily} models and the additional physics included to model the TDE statistics. We then describe {\phaseflow} and how it is linked to {\lgbh} to assign TDE rates to MBHs. 

\subsection{The L-Galaxies semi-analytic models: Dark matter merger trees \& baryonic physics}\label{sec:baseline}

The {\lgfamily} semi-analytic model is a well-tested model that tracks the cosmological evolution of the baryonic component of the Universe on top of dark matter merger trees \citep{Croton06, Guo11,Henriques15,Henriques20}. It has been developed on, and is mainly being applied to, the dark matter merger trees of the \emph{Millennium-I} \citep[MS,][]{Springel05} and \emph{Millennium-II} \citep[MSII,][]{Boylan-Kolchin09} cosmological N-body simulations. In this work, we use the merger trees of the MSII which offer a higher mass resolution compared to the MS simulation. Specifically, the MSII has a dark matter particle resolution of $6.89\,{ \times}\, 10^6 h^{-1}\Modot$ in a box size of $100 h^{-1}$Mpc, enabling a good tracing of the cosmological evolution of halos where MBHs of $10^4\,{-}\,10^8\Modot$ are placed. Originally, the MSII was run by using the WMAP1 \& 2dFGRS ``concordance'' $\Lambda$CDM framework \citep{Spergel2003}. However, the version of {\lgfamily} used here applies the \cite{AnguloandWhite2010} methodology to re-scale it to the cosmology of Planck 2018 data release \citep{Planck2020}. This re-scaling modifies by a factor of 0.96 and 1.12 the MSII box size and particle mass, respectively.

Regarding the baryonic component, the current paradigm of galaxy evolution assumes that, as soon as a dark matter halo collapses, an amount of baryons, equal to the baryon fraction multiplied by the halo mass, also collapses within it \citep{WhiteFrenk1991}. During this process, the infalling baryons are heated up and distributed inside the dark matter halo in the form of a diffuse, spherical, and quasi-static hot gas atmosphere. With time, this gas is allowed to cool down and migrate towards the center of the halo at a rate that depends on redshift and halo mass \citep{WhiteandRees1978,SutherlandDopita1993}. Due to angular momentum conservation, the cooled gas settles in a disk-like structure characterized by a radially exponential distribution. Once the disk becomes massive enough, star formation is triggered giving rise to a stellar component distributed in a disk with specific angular momentum inherited from the cold gas \citep{Croton06}. Right after any star formation event, massive and short-lived stars explode polluting the interstellar medium and injecting energy in their environment, which can warm up and/or eject part of the cold gas of the disk. As a consequence of the ongoing stellar disk growth, some galaxies are prone to become unstable, with the subsequent disk instabilities leading to the formation of a stellar bulge component \citep{Efstathio1982}. Besides supernova explosions, {\lgfamily} assumes that MBHs in the center of the galaxy can prevent the gas from cooling in massive galaxies by injecting kinetic energy into the surrounding medium via quiescent gas accretion directly from the hot gas component \citep[dubbed as ``radio mode'' accretion, see ][]{Croton06}, thus hampering the supply of cold gas to a galaxy's disk. On top of secular processes, {\lgfamily} models the interactions between galaxies, occurring after a given time of the fusion of their parent DM halos. Such interactions include major and minor galaxy mergers and alter the structure of the remnant galaxy by triggering bursts of star formation and leading to the formation of a stellar bulge or pure elliptical structure. Finally, {\lgfamily} also takes into account environmental processes such as ram pressure stripping or galaxy disruptions in its galaxy formation paradigm \citep{Henriques15}.

To improve the time resolution offered by the outputs of the MSII simulation,  {\lgfamily} does an internal time interpolation between two consecutive snapshots, with the time resolution being $dt_{\rm step} \,{\sim}\, 5\,{-}\, 20$ Myr, depending on redshift. 

\subsubsection{Massive Black Holes in L-Galaxies}\label{sec:bhlgbh}

The version of {\lgfamily} used in this work, {\lgbh}, is based on the one presented in \cite{Henriques15} with the modifications included in \cite{IV19,IV20,IV22} and \cite{Spinoso23} to incorporate detailed massive black hole physics.\par
In brief, with respect to the model presented in \cite{Henriques15}, this new version adds a detailed model for the assembly (mass and spin) of nuclear MBHs, the dynamical evolution of MBH binaries, and the production of wandering MBHs \citep[see][]{IV20,IV22}. Concerning the genesis of the first MBHs, {\lgbh} models the spatial variations of the intergalactic metallicity and the Lyman-Werner background\footnote{The Lyman-Werner (LW) band is a specific energy-interval (i.e. $h\nu=[11.2\,-\,13.6]{\rm eV}$) in the UV spectrum. LW photons are responsible for the photo-dissociation of molecular hydrogen \citep[e.g.][]{haiman_rees_loeb1997}.} produced by star formation events to account for the formation of \textit{heavy} (i.e. $M_{\rm seed}\,{\sim}\,10^5 \Modot$) and \textit{intermediate-mass} ($M_{\rm seed}\,{\sim}\,10^{3}\,{-}\,10^{4} \Modot$) MBH-seeds, respectively via the collapse of pristine massive gas clouds and runaway stellar mergers within dense high-redshift\footnote{not to be confused with the ones introduced later in this work, for which we follow a different treatment} NSCs. The formation of \textit{light seeds} ($M_{\rm seed}\,{\sim}\,10^2 \Modot$) after the explosion of the first metal-free stars (PopIII) is accounted for by grafting/inheriting the evolved counterparts of light-seed modeled self-consistently by the GametQSO/dust model \citep[see e.g.][]{Valiante16}. We note that concerning the black hole seeding model presented in \cite{Spinoso23}, we adopt a slightly higher amplitude of the ``grafting probability'', setting the parameter $\rm G_p=0.25$ \citep[see][for the definition of this parameter]{Spinoso23}. This choice is motivated by the normalization of the $z\,{=}\,0$ black hole mass function in the current work.\par
Once the first MBH seeds are formed, galaxy mergers and disk instabilities funnel new gas to the galactic nuclei, making it available for the growth of nuclear MBHs.
%These events fill a gas reservoir around the MBHs (with an efficiency that grows with redshift\footnote{This redshift dependence is motivated by the fact that high-$z$ galaxies are more compact and denser, thus accretion is more efficient at high $z$.}
The gas reaching the center is progressively consumed by the MBH first in an Eddington limited phase, followed by a sub-Eddington one \citep{Bonoli09, IV20}. Episodes of gas accretion, on top of triggering BH growth, also modify its spin \citep{IV20}. The dynamical evolution of massive black hole binaries in a post-merger galaxy is two-fold \citep{Begelman80}. During the pairing phase, the MBH(s) from the satellite galaxy migrate towards the galactic center via dynamical friction \citep[following ][]{BT87}, forming a hard binary upon arrival at the nucleus. Then the hardening phase, where interactions with a circumbinary disc (gas-torque model by \citealp{Dotti15} and preferential growth as in \citealp{Duffell20}) or the surrounding stars \citep[following the model by][]{Quinlan97,Sesana15} assist the gravitational-wave-driven evolution, along potential triplet interactions in the cases that a third MBHs comes in during the inspiral \citep[modeled based on the results of][]{Bonetti18b}. The latter, along with gravitational recoil after merging of MBHs \citep[as described in][]{Lousto12}, can result in wandering MBHs. While {\lgbh} can track the evolution of wandering MBHs, we do not include that in this work for simplicity (see the discussion on TDEs from wandering MBHs in Sect.~\ref{sec:lim_and_prosp}).\\

\begin{figure}
\centering
\includegraphics[trim={0cm 0cm 0cm 0cm},clip,width=0.48\textwidth]{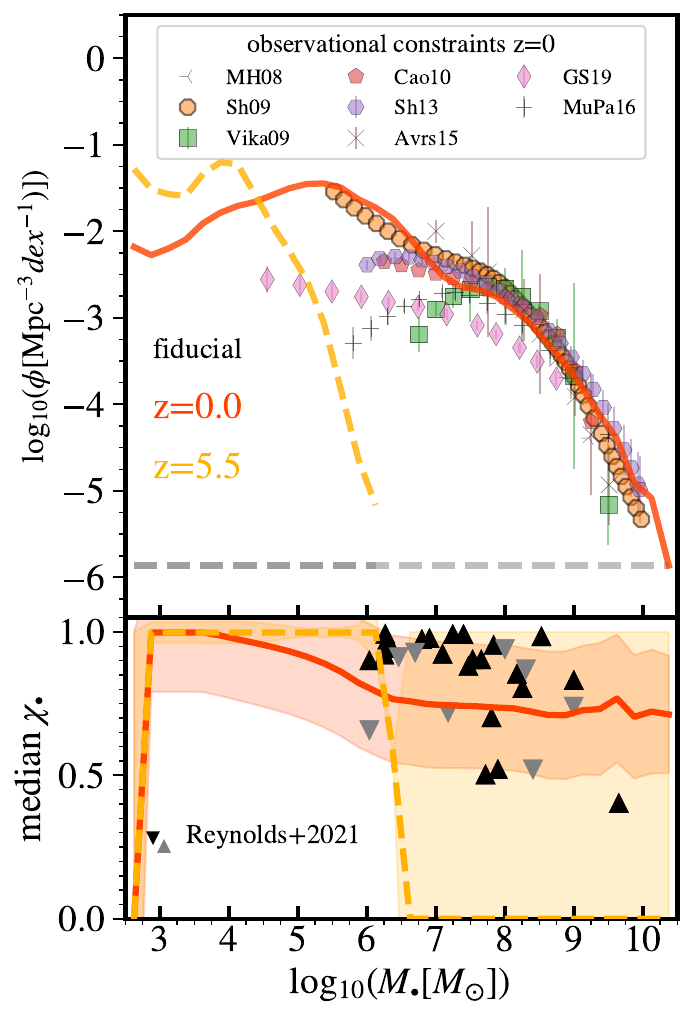}
\caption{MBH mass function (top) and median spin for X-ray bright MBHs (bottom) as a function of the MBH mass $M_\bullet$ predicted by the {\lgbh} model used in this work; data are shown for $z\,{=}\,0$ (red solid line) and $z\,{=}\,5.5$ (yellow dashed line), with shaded areas in the bottom panel referring to the 1$\sigma$ dispersion at a given mass range. In the top panel, the grey dashed line corresponds to the MBH mass function value equal to $1 \mathrm{dex}^{-1}$ per MSII simulation volume; the results are compared with observational data at $z\,{=}\,0$: MH08, Vika09, Sh09, Cao10, Sh13, Arvs15, GS19, MuPa16 refer to the model-dependent constraints on the MBH mass function derived respectively by \citet{MH08},\citet{Cao10},\citet{Gallo19},\citet{Shankar09},\citet{Shankar13},\citet{Mutlu-Pakdil16},\citet{Vika09},\citet{Aversa15}. For spin constraints, we display the upper and lower limits from X-ray reflection spectroscopy \citep{Reynolds21}. For a closer comparison to observational results, the average spin values shown here are for MBHs with a predicted hard X-ray luminosity of  $\log\,L_{HX} \,{>}\, 40$ erg s$^{-1}$.}
\label{fig:lgbh-base}
\end{figure}

As an example of the capability of {\lgbh} to produce a realistic population of MBHs, in Fig.~\ref{fig:lgbh-base} we show the MBH mass function $\phi(M_\bullet)$, where $M_\bullet$ is the black hole mass (throughout this work), and the MBH median spin distribution $\Tilde{\chi}_\bullet(M_\bullet)$ at $z\,{=}\,0$ and $z\,{=}\,5.5$ for the version of {\lgbh} adopted in the current work. We compare our results with available data. Regarding the black hole mass function, as noted by \cite{Shankar19}, all observationally-derived values seem to converge at the high mass-end \citep[$M_\bullet\,{>}\,10^{7.5}\Modot$][]{MH08,Cao10,Gallo19,Shankar09,Shankar13,Mutlu-Pakdil16,Vika09,Aversa15}.  However, constraints in the intermediate-mass range $M_\bullet \,{\sim}\, 10^5 - 10^6\Modot$ are much less stringent. {\lgbh} agrees with the most optimistic estimates from \citet{Shankar09} and over-estimates with respect to all other available constraints. Regarding the spin distribution, the model agrees with the constraints from \citet{Reynolds21} at high masses, yet it predicts high-to-maximal spin for MBHs in the intermediate mass range, where there are no observational constraints. TDEs can potentially offer as a new probe for both the MBH mass and spin distribution in this range (see discussion in Sect.~\ref{sec:Spin_Implications}).%As shown, the model is in good agreement with observations. We also refer to Fig.~\ref{fig:comptwomodels} to show the accuracy of the model to constrain the occupation fraction of MBHs in dwarf galaxies.\\

\subsubsection{The Stellar Environment of Massive Black Holes}\label{sec:gencon}
As mentioned above, the novelty of this work is the inclusion of TDEs within a full galaxy evolution context. To encompass this ambitious task it is necessary to model the stellar environment in which MBHs are embedded, on top of their formation and evolution. In this section, we describe how disks, bulges, and NSCs are included in {\lgbh}. Together with black hole masses, the properties of the nuclear stellar environment will be the input for predicting time-evolving TDE rates with \textit{PhaseFlow}.

\subsubsection*{Bulge and Disk profiles in L-Galaxies}\label{sec:bulges}
Bulges in {\lgbh} grow after galaxy interactions (major and minor mergers) and disk instabilities occurring in massive stellar disks. The specific properties of these events fully determine the final mass and extension (i.e. effective size) of the bulge. We stress that the scale length for disks $R_{\rm eff,d}$ and the effective size of bulges $R_{\rm eff,b}$ (equivalent to the scale radius of a S\'ersic profile) are computed self-consistently inside {\lgbh} by tracing the spin evolution of the galaxy components and applying energy conservation during mergers and disk instabilities. The profile of each bulge is assumed to follow a S\'ersic model, whose steepness (i.e. S\'ersic index) is associated with each bulge by using the observational results of \cite{Gadotti09}, approximately a Gaussian distribution peaking at S\'ersic index $n_s\,=\,3-4$, as implemented in \citet{IV19}. As already mentioned, galactic disks arise as a consequence of gas cooling and star formation events occurring in the center or dark matter halos. Together with galaxy encounters, these events determine the extent of the disk radial profile. Taking this into account, in this work we assume that pure disks with no bulge are characterized by the S\'ersic index $n_s\,{=} \,1$, regardless of redshift, and a scale radius of $R_{\rm gal} \,{=} \, 1.68\,R_{\rm eff,d}$. For the rest of this work, we refer to the sum of the disk and bulge mass as galaxy stellar mass $M_\ast$ (the halo and intracluster stellar mass are neglected during our analysis), while $M_{\rm gal}$ is reserved for the integrated mass of a S\'ersic profile (either a bulge or a disk) with radius $R_{\rm gal}$.
As we will see later, TDE events due to encounters between the nuclear MBHs and stars belonging to the bulge or disk component will be calculated assuming that these density profiles extend all the way to the center of the galaxy.

\subsubsection*{Nuclear Star Clusters in L-Galaxies}\label{sec:condNSC}

NSCs observed in the centers of a great fraction of dwarf and massive galaxies in the local Universe are the densest stellar structures known  \citep[see][for a review]{Neumayer2020}. This inevitably suggests that they might be an ideal nursery for TDEs. In the following paragraphs, we describe a simple \textit{phenomenological model} that we are introducing in {\lgbh} to incorporate NSCs in galaxies (``nucleation''). An extensive and self-consistent model of the birth and evolution of NSCs will be presented in a future paper (Hoyer et al.\ in prep). 

\paragraph*{\bf NSC Mass:}
The mass of an NSC, $M_{\rm NSC}$, is a fundamental property to be determined. To this end, we connect the NSC mass with the total galaxy stellar mass $M_\ast$ of the host system via the following relation derived from observations of clusters in the local universe:
\begin{equation}\label{eq:mnscscale}
  \log_{10}(M_{\rm NSC}/M_\odot)= A + B \log_{10}(M_\ast/10^{9.4}M_\odot)
\end{equation}
with  
\begin{align*}
  A=6.684\;\;\&\;\;B =\begin{cases}
   0.94 , & {\rm if\, } M_\ast>10^{9.4} M_\odot\\
   0.55 , & {\rm if\, } M_\ast\leq 10^{9.4} M_\odot
  \end{cases}.
%\label{eq:const_mnscscale}
\end{align*}

The high mass end of this relation is obtained from the work of \citet{Pechetti20}, while the lower mass end is adapted to the results from \citet{Hoyer23}. We also introduce a $0.5  \, \mathrm{dex}$ uniform scatter to these median values, comparable to the scatter of $0.23  \, \mathrm{dex}$ measured by \citet{Pechetti20} to their relation as well as to the uncertainty on the assumed mass-to-light ratio of about $0.3 \, \mathrm{dex}$ \citep{Roediger15}.

To avoid the formation of too many small clusters, when applying the above mass scaling relation at arbitrarily low galaxy stellar masses and at all redshifts we impose a minimum mass limit $M_{\ast,\rm NSC}^{\rm min} = {5}\,{\times}\,M_{\rm jeans}(z)$ where $M_{\rm jeans}(z)$ is the redshift-dependent Jeans mass for cold gas in the absence of a heat bath \citep{Rees77}. To avoid unphysically massive NSCs, we also impose a maximum mass limit of $M_{\ast,\rm NSC\,\max}$ equal to 95\% that of the galaxy component (bulge, or disk in the absence of bulge) $M_{\rm gal}$. These factors are arbitrary and are kept fixed throughout this work.

\paragraph*{\bf Nucleation:} In this work, we make the simple assumption that only galaxies with an MBH can host an NSC, although we are fully aware that the frequency of co-existence of MBHs and NSCs is observationally not yet fully established, with only a handful of NSCs and MBHs being detected in the same galaxy \citep{Seth08,Graham09,Neumayer12,Georgiev16,Nguyen18,Kimbrell21,Nguyen22,Ashok23,Thater23}.\footnote{Notice the existence of a few nearby galaxies hosting an NSC but lacking an SMBH: M{\,}33 \citep{Gebhardt01,Merritt01}, M{\,}110 \citep{Valluri2005}, or NGC{\,}7793 \citep{Neumayer12}.}
In our toy model, galaxies hosting an MBH can also host an NSC depending on a simple step-function probability:
\begin{equation}
P(M_\ast,z)  = P_0 ,\;\;{\rm for}\;\; M_{\ast,\rm NSC}^{\rm min} <M_\ast< M_{\ast,\rm NSC\,cut-off}
\label{eq:theta}
\end{equation}
where $M_{\ast,\rm NSC\,cut-off}$ is a cut-off mass and $P_0$ is a free parameter. For our {\it fiducial} model, we assume $P_0\,=\,1$. For the case of $M_{\ast,\rm NSC\,cut-off}$ our {\it fiducial} model uses the value $10^{9.75}\Modot$ motivated by the theoretical work of \citet{Antonini15} and the observations of the Local Volume and close galaxy clusters \citep[see e.g.,][]{Hoyer21}. 
 %%DEPRICATED: but we will discuss how results are affected for a lower probability value (see the discussion in Sect.~\ref{sec:discussion}) aka hierarchical model

After formation, we assume that NSCs do not change in mass if the galaxy is evolving secularly (see discussion in Sect.~\ref{sec:reset}) or experiences only minor mergers (in this case, the NSC of the central galaxy acquires the NSC of the merged satellite, following the dynamics of the companion MBH). Indeed, NSCs are extremely compact stellar systems and are expected to be difficult to be destroyed from external tidal fields during mergers \citep[see the works of ][for the detection of stripped nuclei]{Bassino94,Pfeffer14,Mayes21}.

\begin{figure}
\centering
\includegraphics[width=0.48\textwidth]{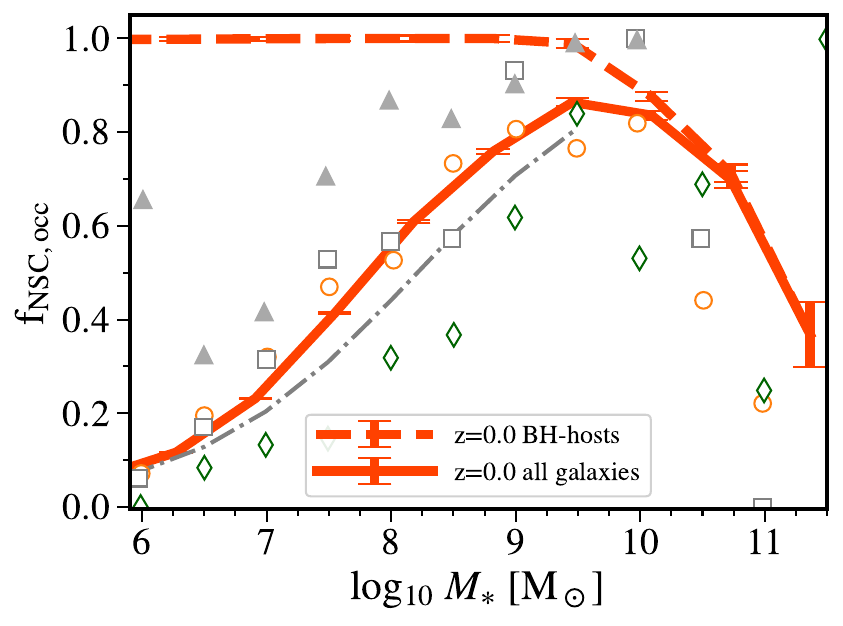}
\caption{The NSC occupation fraction for our {\it fiducial} model as a function of galaxy stellar mass, for all galaxies (solid line) and all galaxies hosting an MBH (dashed line). All $M_\ast\,{<}\, 10^{9}\Modot$ galaxies hosting an MBH have also a 100\% probability of hosting an NSC at creation. The data represents the NSC occupation fraction for the Virgo (orange circles), Fornax (white squares), Coma (grey triangles) clusters, and the Local Volume (green rhombuses) as presented in \citet[][]{Hoyer21}. Our model fits the logistic function for NSC occupation at $M_\ast \,{<}\,10^{9.5} \Modot$ of the same work (thin grey line). We stress that this agreement occurs naturally from the occupation of MBHs per galaxy (see discussion in Sect.\ref{sec:implMBHNSC}).}
\label{fig:lgbh-nucleation}
\end{figure}

Following these assumptions, the model predictions for the NSC occupation fraction at $z\,{=}\,0$ for all galaxies hosting an MBH and for the full population are shown in Fig.~\ref{fig:lgbh-nucleation}. Regarding galaxies hosting an MBH, we see that below the cut-off galaxy stellar mass $M_{\ast,\rm NSC\,cut-off}$, all galaxies also host an NSC, by construction. Above the cut-off mass,  NSCs are present only in galaxies that were smaller at the time of NSC formation and then evolved secularly in stellar mass (see discussion in Sect.~\ref{sec:reset}). When looking at the occupation fraction of the full galaxy population, instead, we see that dwarf galaxies are less likely to host an NSC because of the lower MBH occupation fraction. This is what is driving the total NSC occupation fraction to increase with galaxy stellar mass until approximately the cut-off mass, naturally following the logistic function that is fitted to the observations of \cite{Hoyer21} up to $M_\ast=10^{9.5}\Modot$.

\subsection{From galaxy properties to time-dependent TDE rates with PhaseFlow}\label{sec:phaseflow}

Black hole masses and the properties of the nuclear stellar component of galaxies, modeled in {\lgbh} as described above, provide the starting point for the calculation of time-dependent TDE rates.

The main, ubiquitous generation mechanism for TDEs is thought to be two-body relaxation \citep{Chandrasekhar42,Frank76,Binney08}. In simple terms, stars in the vicinity of an MBH deflect each other's orbits so that a number of them may eventually reach a very small pericentre, closer than its tidal disruption radius, defined as
\begin{equation}\label{eq:r_t}
 r_{\rm t}\approx \left(\frac{M_\bullet}{m_\star}\right)^{1/3} r_\star
\end{equation}
and be disrupted by the MBH of mass $M_\bullet$; here $m_\star$ is the stellar mass and $r_\star$ its radius. The occurrence of these events can be described in terms of the \textit{loss cone theory} (see e.g.,   \citealt{Merritt13,Stone20}), adopting a Fokker-Planck approach to treat two-body relaxation. In particular, in the present work, we make use of the {\phaseflow} code \citep{Vasiliev17}, which is part of the \texttt{AGAMA} toolkit \citep{Vasiliev19}.
{\phaseflow} evolves in time a phase space density profile assuming a spherical and isotropic distribution function $f(E)$, by solving the Poisson and orbit-averaged Fokker–Planck 1-D equations for the stellar distribution function, its gravitational potential, and its density. In general, two-body relaxation would induce an additional, explicit dependence of the distribution function on the angular momentum $J$ of the orbit, which allows for the computation of the rate of stars being tidally disrupted. The rate of stars with energy $E$ whose pericentre becomes smaller than the tidal disruption radius $r_t$ can be computed as the rate of stars whose angular momentum becomes smaller than $\sqrt{2\, G\, M_\bullet\,r_t}$. At each energy, {\phaseflow} assumes the steady profile in angular momentum arising from relaxation \citep{Cohn78}, and directly associates the isotropic profile $f(E)$ to a (per energy) TDE rate. This results in an extra sink term in the energy-only Fokker-Planck equation and a growth term for the MBH mass. {\phaseflow} has been extensively used in the framework of inferring TDE rates, including addressing the impact of the stellar mass function on TDE rates \citep{Bortolas22} as well as predicting realistically partial disruption event rates \citep{Bortolas23}.

\subsubsection{PhaseFlow set-up}
In our implementation, we assume that the stellar system surrounding the MBH is composed by a bi-chromatic population of stars made up of main-sequence stars of 0.38 M$_\odot$ (encompassing $\approx 93$\% of the total stellar mass) and $16 \Modot$ stellar black holes\footnote{The choice for these values for the mass of main sequence stars and stellar black holes comes from the fact that those are close to the average masses for those objects assuming an evolved \citet{Kroupa2001} stellar mass function; in addition, the second moment of the mass function, which sets the relaxation rate of the system and thus the TDE rates, attains a value which is very close to the actual one if we were to assume a complete and evolved stellar mass function (see e.g.\ \citealt{Kochanek16,Stone16,Pfister22,Bortolas22}).} (encompassing the remaining $\approx$7\%). Stars are considered to be destroyed if their separation to the MBH gets below $r_{\rm t}$ (Eq.~\ref{eq:r_t}), where we used $r_\star = 0.44 R_\odot$, which is the expected radius of a 0.38 $\rm M_\odot$ star (see e.g. \citealt{Bortolas22}, Eq.~3, for more details). Once accretion occurs, 30\% of the stellar mass\footnote{At a full disruption, the MBH captures 50\% of the stellar mass. We selected a subsequent mass loss due to radiation and winds removing 40\% of the bound material based on results from observations \citep{Mockler21} and simulations \citep{Bu23}. Simulations including radiative transfer from \cite{Steinberg24} yield a lower percentage of 15\%.} is added to the MBH, while the remainder is assumed to be lost in radiation and the interstellar medium. Stellar black holes are instead captured by the MBH  if they get closer than $8GM_\bullet/c^2$ from the MBH, where $c$ is the speed of light in vacuum and $G$ the gravitational constant. During the evolution of the system, the stellar populations undergo the traditional dynamical phenomena expected in the vicinity to an MBH, i.e. they develop a \citet{BahcallWolf76} cusp, stellar black holes segregate in the center dominating relaxation in the closest vicinity to the MBH and finally, once the system reaches a dynamical equilibrium, expands and lowers its TDE rates as a result of dynamical heating. All these phenomena are captured by {\phaseflow}.

Given the fast performance of the code, we have generated multi-dimensional tables spanning a range of initial central MBH masses and a range of host-environment properties (mass, scale radius, compactness for disks, bulges, and NSCs), encompassing all values predicted by {\lgbh}, as described in what follows. In particular, all runs in the multi-dimensional grid were initially evolved for a Hubble time using 300 bins in the phase-volume, the variable used in the code to parameterize the distribution function \footnote{The phase-volume $h(E)$ is the volume in phase space spanned by all the orbits with energy $E$; unlike the orbital energy it is invariant under adiabatic changes of the potential, like in the case of a central MBH accreting via TDEs, and this makes $h(E)$ a very good choice for this problem.}.

\subsubsection{The PhaseFlow-generated grid of TDE rates}\label{sec:grid}

We model rates depending on their reservoir origin, which could be either a bulge/disk or an NSC. Rates are saved in a large multi-dimensional grid, which includes the parameter space of the MBH, the stellar environment properties, and the time dimension. Rates are \textit{not static} but instead vary with time (not necessarily a steady state is reached). This adds substantial realism to our computation, as many systems dominating the overall TDE rate are often characterized by a very large initial rate which drops by orders of magnitude with time. Assuming the static rate (which would coincide with the $t=0$ rate) thus results in a non-negligible overestimation of TDE rates, especially for the case of NSCs.

\noindent The parameter space mapped is the following:
\begin{description}

\item[{\bf MBH mass:}] {The first input parameter is the central MBH mass. The grid covers the range: 
\begin{equation}
    \log_{10} (M_{\bullet}/M_\odot) \in [2.5,8.0]
\end{equation}
in 34 equally spaced logarithmic steps. Here, the upper limit lies at the high-mass end of the event-horizon suppression defined by the range of values of the Hills mass. The Hills mass is defined as the MBH mass for which the tidal radius is within the horizon radius ($r_{\rm t}<r_{\rm g}$, \citealt{Hills75}). The Hills mass depends on the black hole spin and the infalling star properties and its orbit \citep{Kesden12,Mummery23c}, but it is in the range $M_\bullet = 10^7\,{-}\,10^8\Modot$ for an $0.38 \Modot$ star} (see also  Sect.~\ref{sec:postproc}).\\

\item[\textbf{Galaxy stellar component:}]{To predict TDE rates, {\phaseflow} needs the galaxy stellar mass $M_{\rm gal}$, the scale radius $R_{\rm gal}$ and S\'ersic index $n_s$. We map stellar mass values in a broad range around the MBH mass  $M_\bullet$ values, following this scaling:
\begin{equation}\label{eq:mstarrange}
    \log_{10} (M_{\rm gal}/M_{\bullet}) \in [1,4]
\end{equation}
in 16 equally spaced logarithmic bins. The scale radius can instead vary in the range:
\begin{equation}
    \log_{10}(R_{\rm gal}/R_{\rm scl}) \in [-2,1]
\end{equation}
with  13 equally spaced logarithmic bins, where $R_{\rm scl}$ depends on the stellar mass as \citep{Shen03}:
\begin{equation}
    \log_{10} (R_{\rm scl}/{\rm kpc}) = 0.14\log_{10}(M_{\rm gal}/M_\odot)-1.21 \; .
\label{eq:rgal}
\end{equation}
Finally, for the S\'ersic index we assume all integer values between one and seven, which is the range we probe in {\lgbh}, as described in Sect.~\ref{sec:bulges}. This gives a grand total of $34 \times 16 \times 13 \times 7\, =\, 49{\,}504$ combinations of parameters for the galaxy stellar component.
For galaxies with both a disk and bulge component, the contribution of the disk to TDE rates is ignored, as it is generally significantly lower. Note, however, that small galaxies in {\lgbh} are often pure disks, and their contribution to the total volumetric rate is non-negligible. 
}\\

\item[{\bf NSCs: }]{To estimate the TDE rates in an NSC environment with {\phaseflow}, the NSC mass $M_{\rm NSC}$, effective radius $R_{\rm NSC, eff}$  and density profile are needed. Given the mass range for the galaxy component considered in the above grid (Eq.~\ref{eq:mstarrange}), the NSC mass range follows by applying to this the scaling relation presented in Eq.~\ref{eq:mnscscale} (from combining the works of \citealt{Pechetti20} and \citealt{Hoyer23}). From \citet{Pechetti20} we also use the definition of the NSC effective radius, which correlates with the galaxy stellar mass $M_\ast$ as: 
\begin{equation}
    \log_{10}(R_{\rm NSC, eff}/{\rm pc}) = 0.53+0.29\log_{10}(M_\ast/10^9M_\odot).
\label{eq:rnsc}
\end{equation}
We stress that this relation is especially holding towards the high-mass end but flattens out at low masses \citep{Neumayer2020}. To explore how the results presented in this work depend on the choice of fixed compactness of the NSC, we construct a second grid using $R_{\rm eff, NSC}$ equal to $1/3$ of the radius predicted by the scaling relation. We dub these additional runs as {\it compactNSC}. The {\it fiducial} model follows Eq.~\ref{eq:rnsc} and will be compared with the {\it compactNSC} one in Sect.~\ref{sec:extracomp}.\par
The initial  NSC density profile is assumed to follow the functional forms  of \cite{Saha92} and \cite{Zhao96}: 
\begin{equation}\label{eq:spheroid}
    \rho_{\rm NSC}(r) = \rho_0\left(\frac{r}{a}\right)^{-\gamma}\left[1+\left(\frac{r}{a}\right)^{\alpha}\right]^{\frac{\gamma-\beta}{\alpha}}\exp{\left[\left(-\frac{r}{r_{\rm cut}}\right)^{\xi}\right]}.
\end{equation}
where $\rho_0$ is a normalization constant that ensures the total mass of the cluster is $M_{\rm NSC}$, $a\,=\,R_{\rm NSC, eff}/4.6$ is its scale radius, while $\alpha\,=\,4$, $\beta\,=\,2$, and $\gamma\,=\,0.5$ are respectively the outer, intermediate and inner density slopes. Finally, $r_{\rm cut}\,=\,12a$ is a cutoff radius and $\xi \, =\,2$ represents the cutoff strength. The selected values have been chosen based on the results by \cite{Antonini12}, who explored the formation of NSCs through the infall of star clusters \citep[see e.g.][on the connection between NSCs and globular clusters]{Capuzzo-Dolcetta93,Hartmann11,Arca-Sedda14,Sanchez-Janssen19,Fahrion22,Carlsten22,Leaman22,Hoyer23}. This profile should thus be a valid resemblance to the profile of a recently-formed NSC.\par
Taking into account the scaling relations described above, it is clear that all NSC properties ultimately depend only on the galaxy stellar mass $M_\ast$. Thus, the grid for the TDEs due to NSCs spans only the values of $M_\bullet$ and $M_\ast$, in the same ranges as mentioned for the galaxy stellar component above: total $34\times 16 \, =\, 544$ pairs of parameters. 
}
\end{description}

\begin{figure}
\centering
\includegraphics[width=0.49\textwidth]{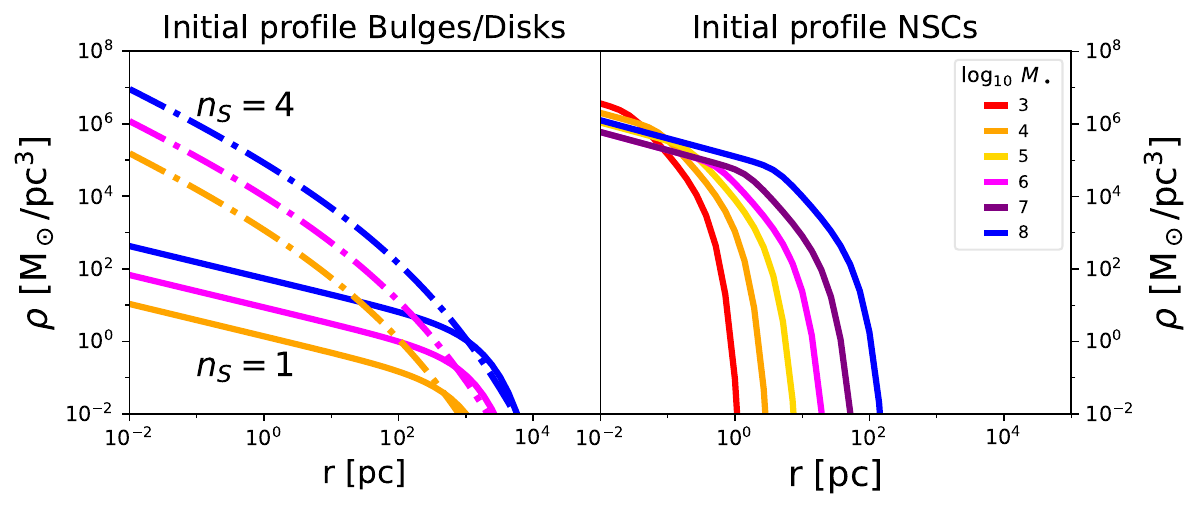}\\
\includegraphics[width=0.45\textwidth]{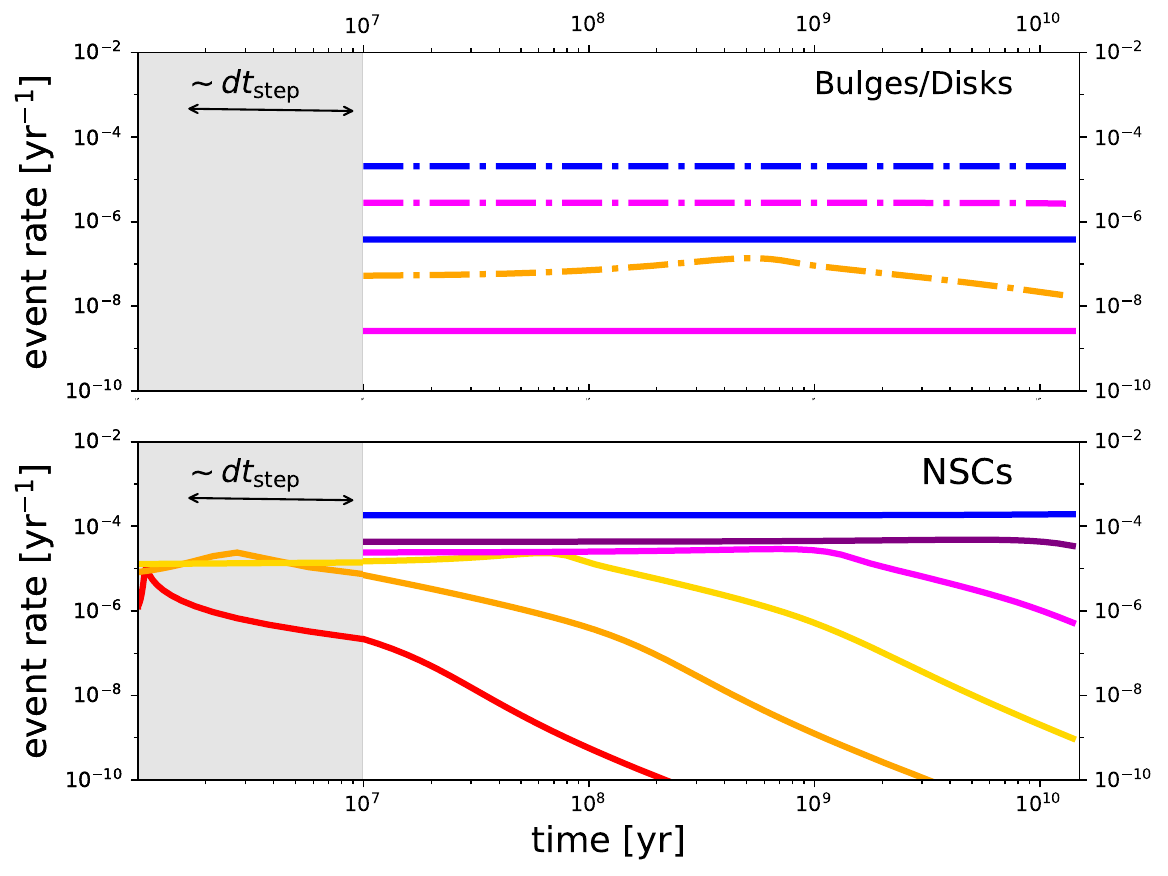}
\caption{{\it Top panels:}  Stellar density profiles for a range of MBH masses as indicated in the inset legend (with the same color-coding applying to all panels of the figures). Disks are assigned $n_S\,=\,1$ S\'ersic profiles (solid lines, left) while bulges $n_S\,=\,4$ profiles  (dotted lines, left). NSCs are instead assigned spheroid profiles from Eq.~\ref{eq:spheroid} (right). The galaxy and NSC host properties scale with $M_\bullet$ as described in the text and are created to be representative of the average environment encountered in {\lgbh}. {\it Middle \& bottom panels:} Tidal disruption event rate evolution with {\phaseflow} when initiating for different MBH mass with the associated profiles from the top panels, displayed separately for bulges/disks (galaxy component, middle panel) and NSCs (bottom). The grey region below $10$Myr indicates the region where we trace unresolved growth and high-rates below the time-resolution $dt_{\rm step}$ for the black holes in NSCs with mass $M_\bullet<10^6\Modot$. This initial phase we refer to as \textit{prompt phase}.}
\label{fig:rates_and_rhos}
\end{figure}

Once the runs with {\phaseflow} are completed and the grid in the parameter space has been fully mapped, we store the resulting event rates, $\Gamma$, which are a function of the aforementioned parameters and time $t$.
For the bulge/disk contribution, the rates are encapsulated in the function  $\Gamma_{\rm gal}(M_{\rm gal},R_{\rm gal},n_s; M_{\bullet},t)$, while for the NSCs, the rates are given by the function $\Gamma_{\rm NSC}(M_{\rm NSC}; M_{\bullet},t)$. We store the rates differently for late times: 
\begin{equation}
    \log_{10} (t /{\rm yr}) \in [7.0,10.146]
\end{equation}
and for early times
\begin{equation}
    \log_{10} (t /{\rm yr}) \in [1.0,7.0]
\end{equation}
with 50 and 60 evenly spaced logarithmic bins, respectively for the two time ranges. We consider separately TDE rates at times below $10^7 { \rm yr}$ given the time resolution of {\lgbh} which spans between $dt_{\rm step} \rm \sim 5\,{-}\,20 \, Myr$, depending on redshift. Specifically, for MBH mass $M_\bullet<10^6\Modot$ the peak TDE rate due to NSCs falls below the time resolution of {\lgbh} $dt_{\rm step}\sim 10$Myr, therefore the initial phase of high TDE rates and growth (dubbed here as \textit{prompt phase}) would not be treated properly by our model (see Fig.~\ref{fig:rates_and_rhos}). To resolve that, when the TDE phase due to NSCs is initialized for these small MBHs, we draw a random time between $1$ and $dt_{\rm step}$ and assign rates from the \textit{prompt phase} tables according to this time. At the same time, we add the integrated mass during the unresolved \textit{prompt phase} to the MBH.\par
In the following section, we outline how we merge {\lgbh} with the $\Gamma_{\rm gal}$ and $\Gamma_{\rm NSC}$ rates constructed from the multi-dimensional grid described above. To guide the reader, Fig.~\ref{fig:rates_and_rhos} shows several examples of the time-evolving TDE rates for a variety of black hole masses and for a fixed set of parameters. In the figure, for each black hole mass, we assume that the corresponding galaxy stellar mass is given from the relation $\log_{10}\,(M_{\bullet}/10^{7.43}{\rm M}_\odot) = 0.62 \log_{10}(M_{\rm gal}/10^{10.5}{\rm M}_\odot)$, which is the best-fit relation we get in {\lgbh} at $z\,{=}\,0$ for MBHs up to ${\sim}\,10^8\Modot$.
The scale radius assigned to the Sérsic profile is here assumed to be an average value of $R_{\rm gal} \,{=} \,R_{\rm scl}$.  The disk corresponds to  $n_s \,{=} \,1$ and the bulge to $n_s \,{=} \,4$. The matching NSC profiles are created by using the scaling relations presented in Eq.~\ref{eq:mnscscale} and Eq.~\ref{eq:rnsc}.\par
As shown from Fig.~\ref{fig:rates_and_rhos}, the main general feature of our model is that the typical, average rates increase with increasing MBH mass, as the stellar reservoirs assigned are increasingly more massive for more massive MBHs. However, there is a strong differentiation between the galaxy component and the NSC component; while bulges and disks profiles often obtain a constant event rate given the large relaxation timescale of these systems, NSCs matching to $M_\bullet \,{<}\,10^6\Modot$ show decay in their initial TDE rate due to their fast relaxation dynamics. Typically, the decay happens earlier for smaller MBHs and smaller reservoirs. Notably, NSCs with $M_{\bullet}$ from $10^4 \Modot$ to $10^6 \Modot$ start with similar TDE rates at $10^7\yrs$, but maintain it shorter times compared to the Hubble time, as the systems quickly reach their equilibrium in the form of a \citet{BahcallWolf76} cusp and subsequently expand due to the dynamical heating resulting from efficient relaxation. That proves the importance of including the analysis of the initial rates that fall below {\lgbh} time resolution.

\subsection{TDE rates in L-Galaxies} \label{sec:starfeed}

\begin{figure*}
\centering
\includegraphics[width=0.9\textwidth]{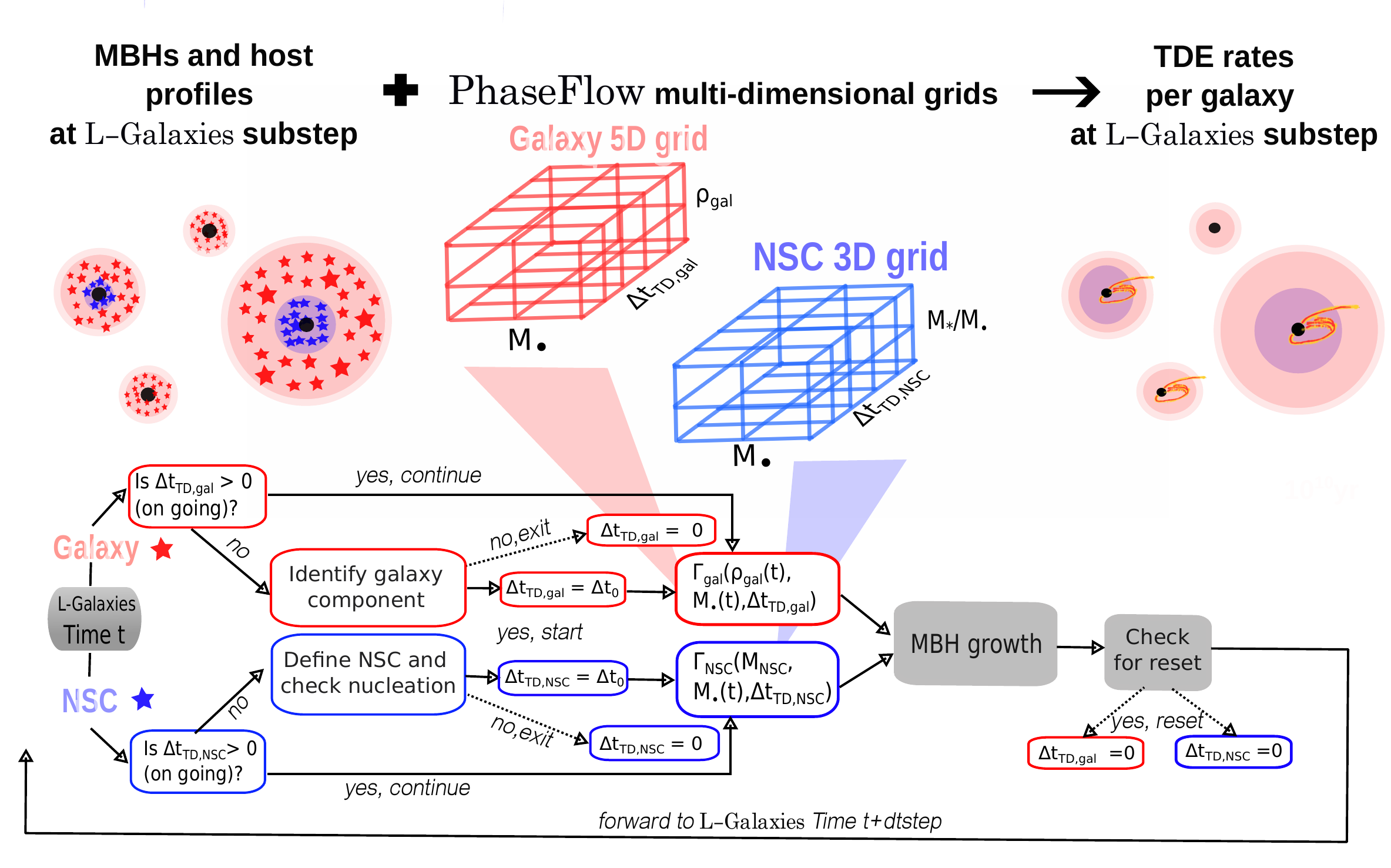}
\caption{Flowchart of the current scheme for estimating TDEs within a single time-step of {\lgbh}. The decision-making tree is colored red and blue for the galaxy (bulge or disk in the absence of bulge) and NSC components, respectively. With the density profile $\rho_{\rm gal}$ we refer to all galaxy component parameters $\,\left\lbrace M_{\rm gal}, R_{\rm gal}, n_S \right\rbrace$ that are used to initialize the S\'ersic profile. TDE rates from the galaxy as a whole and its NSC are effectively treated independently (each process has its clock $\Delta t_{\rm TD}$) and can be active at the same time, while they sum together to produce the instantaneous rate of each MBH.}
\label{fig:flowchart}
\end{figure*}

The final step consists of joining in a self-consistent way the TDE rates calculated by  {\phaseflow} and the properties and stellar environments of MBH predicted by {\lgbh}. We schematically show this in Fig.~\ref{fig:flowchart}, and we describe the details below. 

\subsubsection{Event rates and mass accretion}

As soon as an MBH forms inside a galaxy in {\lgbh}, we set a TDE rate depending on the existence of a stellar bulge/disk and/or an NSC. Effectively, we look up the multi-dimensional grid generated with {\phaseflow} for the rate $\Gamma_i$, with $i \, {=} \, {\rm gal}$ for the galaxy component and $i \,{=} \,{\rm NSC}$ for the NSC. From this, we derive the time-evolving rates for each MBH and stellar environment. For galaxy and NSC properties we look up for the closest values in the grid, while we interpolate in the MBH and time dimension.\footnote{In only a few instances values predicted by {\lgbh} are outside the parameter space covered by the grid. In these rare cases, we use the closest valid value.}\par

When TDEs start being produced within a galaxy, either from the relaxation of the bulge/disk or the NSC, we start a clock, $\Delta t_{\rm TD,i}$. In the successive steps of the {\lgbh} run, the {\phaseflow} grid is checked at the corresponding subsequent times. As described in the next section, we reset this clock to zero only when the MBH and/or the galaxy have substantially changed, which can happen, for example, after mergers.\\

\noindent Next, we define the fraction of the stellar mass accreted by the MBH after it has been tidally disrupted:\\
$f_{\ast,\rm TDE}= \left\{\begin{array}{l}
0.3,\; {\rm if }\; M_{\bullet} \le 10^8{\rm M}_\odot\\
 1,\;{\rm otherwise }
\end{array}
\right. .$\\
In reality, around the Hills mass 
  both direct captures and disruptions could take place, depending on the orbit of the infalling star \footnote{as mentioned earlier, $M_\bullet \,{=}\,10^8\Modot$ is the approximate mass limit for event-horizon suppression; beyond this MBH mass stars are not disrupted, therefore $f_{\ast ,\, \mathrm{TDE}} = 1$ instead of $0.3$.}, but we don't expect this to change our results both on the TDE rates and the BH growth which, in this mass range, is anyway dominated by gas growth \footnote{For the TDE rate $10^{-5}$yr$^{-1}$ a $10^8 \Modot$ will grow over $10$Gyr only $4\times 10^{-4}$ its mass, approximately equal to the gas growth at $10^{-3}$ of the Eddington-limited accretion episode over $15$Myr.}. \par
Provided the TDE rate and the accretion fraction $f_{\ast ,\, \mathrm{TDE}}$ per event, the mass accreted by the central MBH from each stellar component\footnote{Note that we consider an NSC as a decomposed part of the central bulge/disk, that we subtract separately from.} during each time-step of {\lgbh} is given by:
\begin{equation}
  d  M_{i,\bullet-\mathrm{acc}}= dt_{\rm step} \, \Gamma_{i} \,f_{\ast,\rm TDE}\,m_\ast 
\label{eq:dmbh}
\end{equation}
% \begin{equation}
%   d  M_{\bullet,\mathrm{acc}}= dt_{\rm step} \sum_{i}^{\rm gal,NSC} \, \Gamma_{i} \,f_{\ast,\rm TDE}\,m_\ast 
% \label{eq:dmbh}
% \end{equation}
and the mass subtracted ($i=\rm gal$ or $i=\rm NSC$) is:
\begin{equation}
   d M_{i, \mathrm{loss}}=  dt_{\rm step} \left[ \, \Gamma_{i}\,(+\, \Gamma_{\rm NSC} \;{\rm if}\,i=\rm gal) \, \right] \,m_\ast .
\label{eq:dMstars}
\end{equation}
Notice that, if both a bulge/disk and an NSC are present, we sum the two TDE rate contributions. The results of the mass growth of MBHs through this mechanism are discussed briefly in Sect.~\ref{sec:growth}.

%We stress that Equation \ref{eq:dmbh} allows the rate of events to be so high that intermediate-mass MBHs grow with average rates higher than their Eddington limit. This is important for the lowest seed masses that might be assigned rates that allow super-Eddington accretion. However, numerically and for the current set of parameters, the time resolution of L-Galaxies {\lgbh} does not allow for an average super-Eddington accretion. 

\subsubsection{Conditions for changing/resetting TDE rates}\label{sec:reset}

As mentioned above, {\phaseflow} is able to capture the time evolution of the stellar density profile as a result of relaxation and MBH growth due to star accretion. Thus, once TDEs start taking place within a galaxy in {\lgbh}, we follow their time evolution based on the grid provided by {\phaseflow}.

However, dramatic events, such as mergers, disk instabilities, and starbursts, can lead to major changes in the stellar environment surrounding the MBH. In these cases, we \textit{reset} the clock for TDE rates ($\Delta t_{\rm TD,i}$), and the new, unrelaxed stellar system starts again evolving towards a new relaxed state until the next violent dynamical event.

The conditions under which we consider it relevant to check the state of the TDE process taking place around an MBH are the following ones:
\begin{description}
%\item If a small bulge is exhausted (which is rarely the case,% due to $M_{\bullet}$ - $M_\ast$ coevolution) the rates are assigned to the disk.
\item[{\bf Changes in MBH mass:}]  If the mass accreted by the MBH via gas accretion is three times larger than the mass accumulated from TDEs, then the $\Delta t_{\rm TD,NSC}$ clock is reset again and the NSC mass and properties are adapted to the galaxy stellar mass at that time\footnote{This condition guarantees the MBH does not move by more than three mass bins away from its initial mass on the {\phaseflow} grid, at which point the assigned TDE rates we expect to be significantly different}. For bulges and disks, rates are less time-dependent (see discussion in Sec.~\ref{sec:phaseflow}), so we omit this condition.
\item[{\bf Changes in the galaxy component mass:}] If during one  {\lgbh} time-step  the bulge or disk stellar mass increases by more than 20\%,  the clock $\Delta t_{\rm TD, gal}$ is reset. Such variations in stellar mass in a short amount of time is a typical change in post-starburst galaxies \citep{Kaviraj07,vanVelzen18,Wild16,Wild20}. Note that events that do not modify the total stellar mass can also lead to a reset of the rates (e.g., the transfer of mass from the disk to the bulge during disk instabilities). 
\item[{\bf Changes in NSC mass:}] We do not expect all changes of the galaxy component to affect the TDE rate evolution in NSCs, since these can continue relaxing at their own pace unaffected by the changes at larger galactic scales. However, as described in Sect.~\ref{sec:condNSC}, an NSC can undergo significant mass changes following the host galaxy changes. Following the same approach described above, we select that NSC that grow more than ${>}\,$20\% within one {\lgbh} time-step will satisfy the condition of resetting the clock $\Delta t_{\rm TD,NSC}$. %The threshold is selected as the case of bulges so that we minimize the number of free parameters. 
This threshold is selected to be equal to that of the galaxy component in order to minimize the free parameters of the model.\par
\item[{\bf Changes in NSC to MBH mass ratio:}] 
While there are no clear constraints on the limits of the mass ratio between an NSC and the MBH at its center, namely $\mathcal{D}\,=\, M_{\rm NSC} / M_{\bullet}$, we decided to put a lower limit on this ratio so that an NSC gets to be (re)generated only if $\mathcal{D}\,{>}\, \mathcal{D}_{\rm nuc}$. Only few physically-motivated arguments can provide some insight for the allowed value of $\mathcal{D}$: a) MBHs cannot be arbitrarily massive if they form in the cluster, e.g. simulations of runaway stellar collisions by \cite{Kritos23} indicate that $\mathcal{D} \, >\, 1$ and b) the NSC destruction from binary MBHs e.g. total and partial disruption of clusters from intermediate-mass MBH binaries takes place for values $\mathcal{D}\,{<}\,5\;{\rm and}\;<15$ according to \cite{Khan21}. Given the theoretical uncertainties of the NSC-MBH symbiosis and provided local NSCs are observed with even lower values \citep[$\mathcal{D} \, \sim\, 0.01$][]{Neumayer12}, we select conservatively $\mathcal{D}_{\rm nuc}\, =\, 0.1$. We stress that this parameter (along with $M_{\ast,\rm NSC cut-off}$) can be later substituted from a physically motivated formulation when the NSC model is implemented in {\lgbh} by Hoyer et al.\ (in prep). 
\end{description}

We have checked that variations within an order of magnitude of the thresholds mentioned above do not significantly affect the results on the volumetric rates presented in this work. However, we will further discuss some of these assumptions in Sect.~\ref{sec:extracomp}.

\subsubsection{Treatment of TDEs in AGNs}\label{sec:agnmodel}

When MBHs experience phases of gas accretion, they shine as active galactic nuclei (AGN). These phases can be coeval with events of stars disruptions.
The two processes may not be completely unrelated to each other. For instance, it has been proposed that TDE rates can be enhanced due to the alignment of retrograde orbits of NSC stars with the MBH accretion disk \citep{Generozov23,Nasim23} or due to the turn-off of the disk \citep{Wang23c} or due to the quadrupole moment of the disk modifying the loss cone\citep{Kaur24}. However, by construction, our model does not account for such possible rate enhancement since gas physics are not included in {\phaseflow}. 
Regarding the impact of TDEs on the AGN disk, the few TDE interpretations of fast flares in AGN light curves suggest a (temporal) post-flare increase \citep{Blanchard17,Liu20} and others a decrease \citep{Cannizzaro22,Cao23} of the AGN luminosity. Given the high uncertainty, we assume that the disk recovers quickly (compared to the time to the next TDE) or is not interrupted at all by TDEs so that both the TDE and AGN luminosity are kept as independent properties of each MBH. We do not expect this assumption to affect the global rates, due to the small AGN fraction at $z\,{\sim}\,0$ (${<}$10\% of MBHs are active) unless the TDE rates are significantly boosted in the presence of an accretion disk (by a factor of 10 or greater).

\subsubsection{From simulated to observable TDE rates}\label{sec:postproc}

Once TDEs are carefully linked to the MBHs and their galaxies evolving within {\lgbh}, we need to transform the individual rates to observable events to finally compare our predictions with the events picked up by time-domain surveys. Here we keep the following minimum number of assumptions to translate simulated TDE rates to observed ones: 
\begin{itemize}
\item[$\bullet$] We assume 
 that all TDE events are full disruptions. In reality, for massive stars, only those with a pericenter that is a fraction of the tidal radius are fully disrupted (otherwise they are partial), and the energy output of the events may depend on the penetration factor of the star. Moreover, provided that a handful of partial disruption events have been identified \citep{Payne21, Malyali23}, we discuss the impact of such a choice in Sect.~\ref{sec:lim_and_prosp}.
 
\item[$\bullet$] We assume that all TDEs  give a luminosity that falls within the sensitivity of current surveys, i.e., all events are detectable, if there is no AGN contribution. If there is no AGN contribution. In the presence of an AGN, we ignore TDEs in MBHs with AGN bolometric luminosity $L_{\rm bol}\,{>}\,10^{42}$erg/s ({\it fiducial} model) unless noted otherwise. Note also that we are not including any obscuration and line-of-sight effects.

\item[$\bullet$] Regarding the event horizon suppression, the predicted TDE rates for each MBH are considered either visible or not depending on the Hills mass of the black hole itself. The Hills mass depends on (i) the mass $M_\bullet$ and spin $\chi_{\bullet}$ of the MBH, (ii) the age of the loss-cone since the cluster/galaxy component was last reset $\Delta t_{\rm TD,i}$  and (iii) the mass of the infalling star.   For this last point, we draw the stellar mass $m_\ast$ of the disrupted star  from a truncated Kroupa mass function ($m_\ast\,{<}\,1.5\Modot$). Given the values above, each MBH is assigned a Hills mass  $M_{Hills}(m_\ast,\Delta t_{\rm TD,i}, \chi_{\bullet})$ based on the tabulated results of \citet[][Table D1]{Huang22} who have performed a series of simulations of solar-metallicity stars disrupted by MBHs\footnote{These authors may have used slightly different environmental set-ups, but we expect the values to be quantitatively close.}. If $M_{\bullet}\,{>}\,M_{Hills}$, we omit this MBH from the sum of the bulk rates. Otherwise, we assume that all events are full TDEs and have the rate assigned to the MBH.\footnote{From theory, e.g. \citet{MacLeod12} and \cite{Kochanek16}, for a given $M_\bullet$, TDE rates scale with the initial mass function $dN\,{/}dm_\ast$ and power-law $m_\ast^{1/6}$. This will matter mostly beyond the maximum Hills mass for $m_\ast\,{=}\,0.38\Modot$ ($M_\bullet\,{>}\,10^8 \Modot$), a mass range that TDE rates are shown only indicatively. Although the dependence on the initial mass function is captured with the sampling of stars, we do not rescale rates with this power-law dependence for massive monochromatic stars when adding them to the bulk rates.}
\end{itemize}

Note also that, given the mass resolution of the dark matter simulation used and of the multidimensional grid of TDE rates, we omit from the analysis the TDE rates from galaxies with masses $M_\ast\,{<}\,10^{5.5}\Modot$ and/or MBHs $M_\bullet\,{<}\,10^{2.5}\Modot$.
\textit{Finally, note that given all the assumptions above, the volumetric rates presented below can be considered optimistic. }

\section{Results}\label{sec:Results}

In this section, we present our main results  on the predicted TDE rates for the {\it fiducial} choice of parameters and how they compare with the TDE rates observed by the latest time-domain campaigns. We also discuss the implications of our model for the time evolution of TDE rates and the general properties of the MBHs and galaxies hosting TDEs. The interpretation of the results in view of  the population of NSCs and MBHs, the implications for MBH spin and growth through TDEs are  discussed later in Sect.~\ref{sec:discussion}.

\begin{figure*}
\centering
\includegraphics[width=0.485\textwidth]{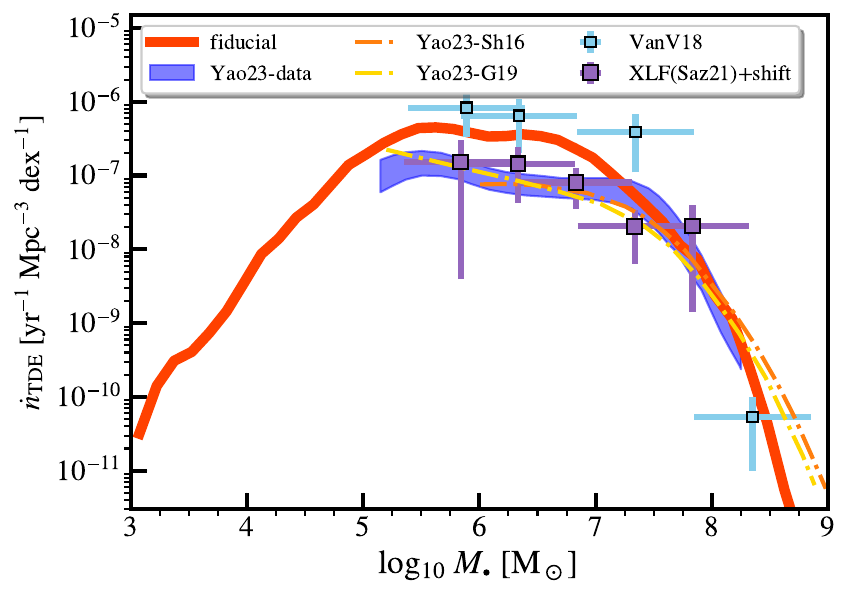}
\includegraphics[width=0.485\textwidth]{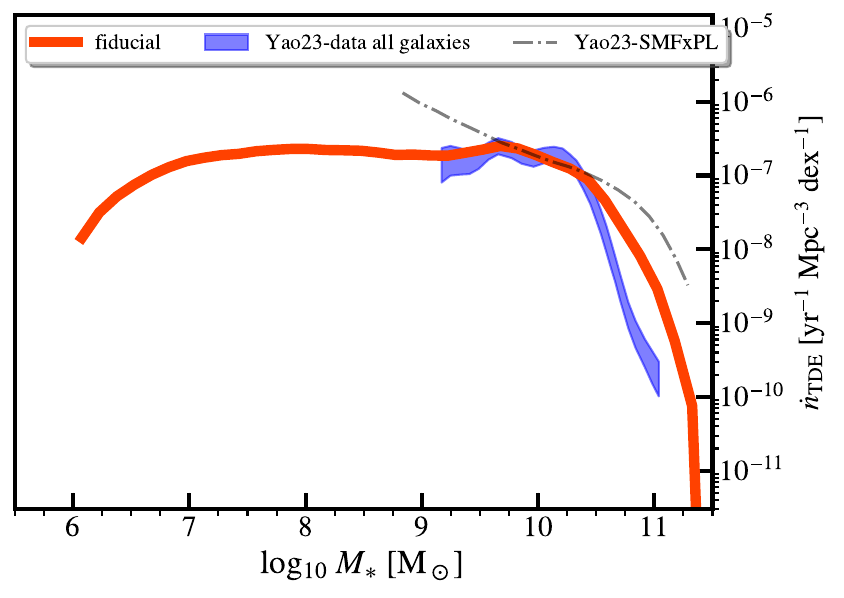}
\caption{Volumetric TDE rates per black hole ($M_{\bullet}$, \textit{left}) and galaxy ($M_\ast$, \textit{right}) log mass at $z\,{=}\,0.0$ for the {\it fiducial} model (solid line). Our model is compared against the constraints (here plotted with the error range) from \citet{Yao23} for all MBHs (\textit{left}) and all galaxy types (\textit{right}). For reference, we display the previous constraints from \citet{vanVelzen18} and the luminosity function from \citet{Sazonov21} transformed to a mass function (see the details in the text). We also display the model lines of \citet{Yao23} for the TDE rates as a function of a) MBH mass, assuming the \citet[][maroon shorter dash-dotted line; Sh16]{Shankar16} and the \citet[][cyan long dash-dotted line; G19]{Gallo19} MBH mass functions, both including the event-horizon suppression b) galaxy stellar mass, obtained by multiplying the galaxy stellar mass function with the power-law dependence of rates on $M_\ast^{-0.41}$ (grey dotted-dashed, SMFxPL).}
\label{fig:globalrates}
\end{figure*}

\subsection{Volumetric TDE rates} \label{sec:VolumetricTDEsResults}

In Fig.~\ref{fig:globalrates}, we present our predictions for the volumetric rates as a function of MBH and galaxy stellar mass for our {\it fiducial} model, comparing them with the recent constraints of \cite{Yao23} derived from a sample of 33 TDEs from the ZTF survey. As shown in the left panel of Fig.~\ref{fig:globalrates}, our model can reproduce the overall shape of the volumetric rates, including the flattening at intermediate MBH masses and the drop for $M_{\bullet}\,{>}\,10^{6.5}\Modot$. \cite{Yao23} also derived a simple model for the volumetric rates as a function of MBH mass using the \citet{Gallo19} and \citet{Shankar16} mass functions and including the event-horizon suppression (dotted-dashed lines in the left panel of the figure), and our predictions are in overall agreement with these models.\par

We note that the higher normalization with respect to \cite{Yao23} at intermediate masses  ($M_\bullet=10^{5.5}-10^{7}\Modot$)  is due to the presence in our model of many black holes in this mass range in dwarf galaxies ($M_*=10^6-10^9\Modot$), 
 which is a stellar mass range still not constrained by current datasets (however see discussion on partial disruption events in Sect.~\ref{sec:discussion}). Also, \cite{Yao23} assign to the detected events black hole masses based on the $M_\bullet$--$\sigma_\ast$ relation from \citep{Kormendy13}, which predicts smaller MBH masses in the low stellar mass regime with respect to the predictions of our model and the observational results of, e.g., \cite{Reines15}. Alternatively, we could obtain a lower normalization in the volumetric rates at intermediate masses by decreasing the MBH and/or the NSC occupation fraction or allowing smaller NSC masses in the model (as discussed in Sect.~\ref{sec:implMBHNSC}). Moving to the low-mass regime ($M_\bullet<10^{5.5} \Modot$), we find that the turnover is due to lower average NSC masses and greater ages around small MBHs (which translates into significantly reduced TDE rates by $z\,{=}\,0$).\par

%It also cannot be attributed to inefficiency or poor observability of disruption events towards the faint-end regime, as proposed, for example, by \citet{Ryu20} and \citet{Wong22}, as we assume that all our modeled events are observable (see Sect.~\ref{sec:postproc}). 
%While this work was being prepared, one of the two dim events contributing to this mass regime, namely {\it AT2020vdq}, was classified as partial disruption due to a rebrightening \citep{Somalwar23a}.

Our results are also broadly in agreement with the volumetric rates as a function of MBH derived from the eROSITA X-ray luminosity function assuming an Eddington-limited accretion and a simple volumetric correction\footnote{computed simply as $M_{\bullet}[M_\odot] \,{=}\, k_{\rm bol}\,L_{X}\,{/}\,L_{\rm edd,1}$ where $L_{\rm edd,1}$ the Eddington luminosity for a solar-mass compact object and $L_{X}$ the soft X-ray luminosity} of $k_{\rm bol}=15$. The two classes of optical-UV and X-ray TDEs might not be eventually different as inferred from recent studies \citep{Guolo24}, so we anticipate the two mass functions to be similar (which is the case within errors with the simple transformation we performed). 
Finally, in the left panel of Fig.~\ref{fig:globalrates} we also plot the previous constraints from \citet{vanVelzen18} for reference. These are higher than both the model predictions and the \cite{Yao23} constraints at all MBH masses below event-horizon suppression. This could be due to the smaller number statistics (17 TDEs) and the heterogeneous nature of the \citet{vanVelzen18} sample \citep[see discussion for the low-end of the luminosity function,][]{Yao23}.

%The inferred total TDE rates is found to be $2.8\times10^{-7}$Mpc$^{-3}$yr$^{-1}$, much lower from the predicted range from first estimates $10^{-6} - 10^{-5}$Mpc$^{-3}$yr$^{-1}$ \citep[see e.g.][]{Wang04}. We attribute this to the fact that we have not used steady-state rates, as it was frequently done during these studies. Secondly, our model includes stellar black holes that mass segregate, enhance the relaxation process, and render the decline of TDEs in time faster. Lastly, unlike many studies, we allowed for a realistic stellar environment in a wide range of galaxy stellar masses (from dwarf to BCG galaxies) generated self-consistently by {\lgbh}. 

In the right panel of Fig.~\ref{fig:globalrates} we present again the model predictions  of  the volumetric rates, but now  as a function of the host-galaxy stellar mass, $M_\ast$. As shown, our predictions at $z\,{=}\,0$ display a broad agreement with the shape of the distribution from \cite{Yao23}.\par 
%We note that the difference in the normalization could be alleviated by increasing the MBH and/or the NSC occupation fraction or allowing larger NSC masses in the model (as discussed in Sect.~\ref{sec:implMBHNSC} \& Sect.~\ref{sec:extracomp}).\par
To describe the distribution of observed TDEs, \cite{Yao23} derive a model using the galaxy stellar mass function coupled with a power-law dependence of the rates on galaxy stellar mass ($M_{\ast}^{-0.4}$, originally from the rate dependence on MBH mass $M_{\bullet}^{B}$, with $B=-0.25$). A similar functional form has been used previously, e.g., for various local galaxies \citep{Wang04}, and for galaxy central cores and cusps \citep{Stone16}. In the range $M_\ast = 10^{9.5}\,{-}\,10^{10.5}\Modot$ our model prediction can be fitted with a similar power-law, but we predict a different functional form at the low mass end.

\begin{figure*}
\centering

%%%VERSION 3 AGAINST RATES FOR DIFFERENT SPLITS FROM LGBH OUTPUT, BUT RESET GAL/NSC generalize to <100Myr ( as \textit{young})
\includegraphics[width=0.49\textwidth]{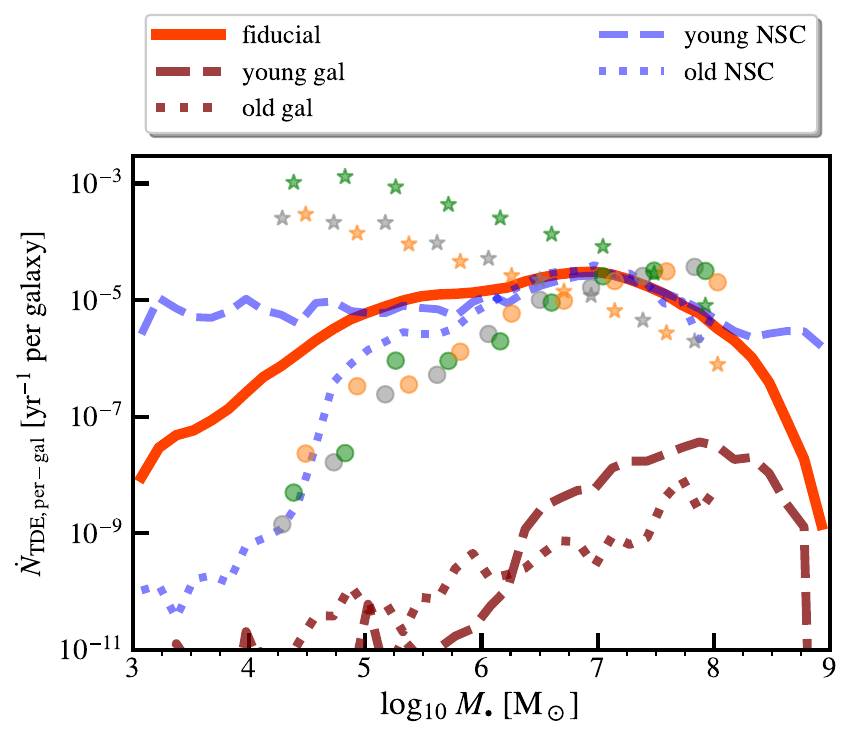}
\includegraphics[width=0.49\textwidth]{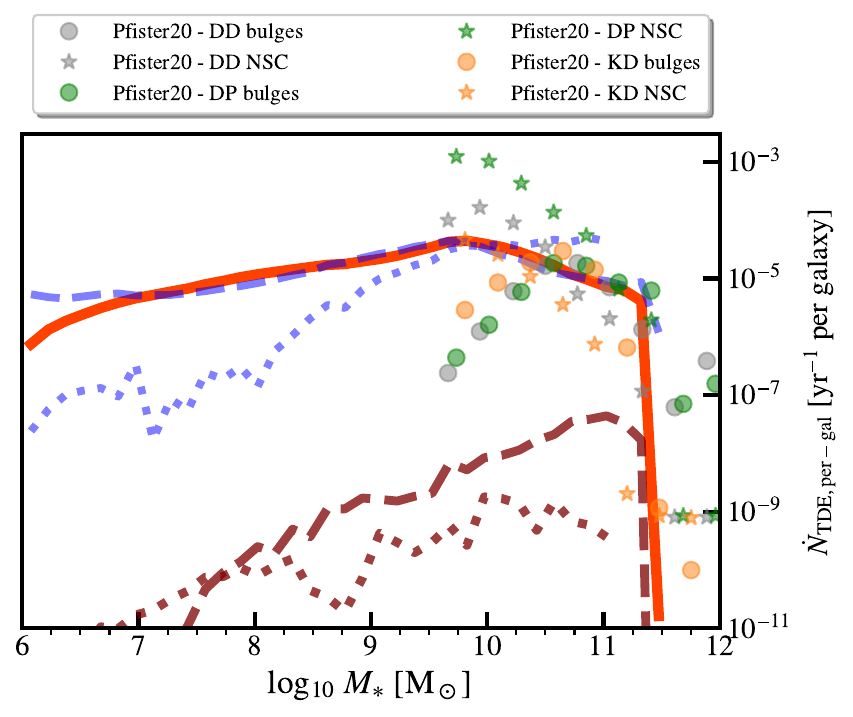}\\
\caption{Average TDE rates per log mass of MBH $M_{\bullet}$ (\textit{left}) and of MBH-host galaxy $M_\ast$ (\textit{right}) for the {\it fiducial} model at $z\,{=}\,0.0$ (solid red line, tagged as ``all''). We average separately over NSC rates (light-blue lines) and galaxy component rates (maroon lines), and subsequently splitting to just restarted (\textit{young} systems with $\Delta t_{\rm TD} \,{<}\, 30\, \mathrm{Myr}$, dashed line) and after a long time (\textit{old} systems with $\Delta t_{\rm TD} \,{>}\, 3 \, \mathrm{Gyr}$, dotted lines). For comparison, we present the results of \citep[][]{Pfister20} for three different scaling relation pairs of MBH-galaxy stellar mass \& NSC mass-size adopted in their work (DD, DP, and KD as defined in Table~2 of \citealp{Pfister20}) to bracket the uncertainties following these hypotheses.}
\label{fig:pergalaxy}
\end{figure*}

%PER GALAXY
\subsection{Per-galaxy TDE rates}

In Fig.~\ref{fig:pergalaxy}, we show the per-galaxy TDE rates as a function of MBH mass (left panel) and stellar mass of galaxies hosting an MBH (right panel). We further split the contributions to the total TDE rates from NSCs and the galaxy component\footnote{As mentioned in the model description, the galaxy component contributing to TDEs is assumed to be the bulge, or the disk in the absence of bulge.}, further subdivided into \textit{old} ($\Delta t_{\rm TD} \,{>}\, 3 \, \mathrm{Gyr}$) and \textit{young} ($\Delta t_{\rm TD} \,{<}\, 30 \, \mathrm{Myr}$) systems. We immediately see that the contribution of NSCs to the total rates dominates across all masses. The  rates from the galaxy component increase both with MBH and galaxy stellar mass up to the event-horizon suppression mass. However, they are on average \emph{more than three orders of magnitude} lower when compared to the contribution from the NSC at fixed MBH/host galaxy stellar mass. This result has important implications for the expected nucleation fraction of galaxies, which we will further discuss in Sect.~\ref{sec:implMBHNSC}. Indeed, based on these results, \emph{TDEs should be predominately observed in the presence of NSCs}. Current observations of TDEs are primarily from distant enough sources where NSCs remain unresolved.\par

%NSC distribution behavior per age
When looking at the difference between \textit{old} and \textit{young} systems, we see a different behavior for the NSCs and the galaxy contributions. Recently formed NSCs give the highest contributions to the per-galaxy rates, with approximately constant values of ${\sim}\, 10^{-5} \, \mathrm{yr}^{-1}$ per MBH/galaxy (see the bottom panel of Fig.~\ref{fig:rates_and_rhos}, where we showed that $M_\bullet = 10^{4}-10^7 \Modot$ have similarly high rates at $t=10^7$yr). Instead, \textit{old} systems below $M_\bullet\,{<}\,10^7 \Modot$ exhibit a power-law dependence on MBH mass, namely $\dot{N}_{\rm per-MBH}\,{\propto}\, M_\bullet^{B}$ with $B\,{=}\,1$ (compared to $B\,{=}\,0$ for \textit{young} systems under the same definition). The \textit{old} NSCs have on average small mass compared to their MBH. For these systems, the relaxation time becomes significantly short, thus the loss cone gets quickly depleted. In the low mass regime ($M_\bullet <10^5 \Modot$), the majority of systems have rates from \textit{old} rather than \textit{young} NSCs and that is why the global per-MBH rates drop towards lower masses. This has specific implications for the dwarf galaxy regime (yet to be probed with observations), where per-MBH/per-galaxy TDE rates are higher by a factor of $>$10 for \textit{young} NSCs over \textit{old} NSCs. Therefore, our work suggests that a promising opportunity of sampling TDEs is selecting recently interacting systems, with a predicted per-galaxy (volumetric) rate of ${\sim}\, 1\,{\times}\,10^{-5}$yr$^{-1}$(${\sim}\,1\times 10^{-7}$yr$^{-1}$Mpc$^{-3}$dex$^{-1}$). This prediction holds down to the least massive dwarfs, as marked by the flattening of the rates as a function of galaxy stellar mass (for both per-galaxy and volumetric TDE rates, see respectively the right panels of Figs. \ref{fig:globalrates} and \ref{fig:pergalaxy}). This revision of rates in dwarf galaxies may have a significant impact on theoretical predictions of TDEs in intermediate-mass MBHs. The galaxy contribution shows a different behavior with age. First of all, we note again that TDEs from the galaxy component do not strongly depend on time and that bulges lead to larger rates compared to pure disks (see discussion of Fig.~\ref{fig:rates_and_rhos} in Sect.~\ref{sec:phaseflow}). 
In Fig.~\ref{fig:pergalaxy}, we see that \textit{old} systems have higher per-galaxy rates with respect to the \textit{young} galactic components at the lower MBH $M_\bullet\,{<}\,10^{6.5}\Modot$ and galaxy  $M_\ast\,{<}\,10^8\Modot$ masses. We attribute this to the morphological differences between \textit{old} and \textit{young} galaxies in this mass range: {\lgbh} predicts that 
\textit{young} systems are mostly pure disks (with an average bulge-to-total mass ratio of $\langle B/T \rangle = 0.075$), while \textit{old} systems are preferentially bulge-dominated ($\langle B/T \rangle = 0.3$). At the high mass end, instead, we see that \textit{young} galaxies have higher per-galaxy rates than \textit{old} ones as a function of galaxy stellar mass. In this case, the difference can be attributed to the differences in the $M_\bullet-M_{\rm gal}$ ratio for \textit{young} and \textit{old} systems: at fixed galaxy stellar mass as an effect of early on-set of MBH growth, \textit{old} systems are preferentially hosting heavier MBHs, whose galaxy-component rates are lower. In both cases the per-galaxy rates follow a power-law of the form $\dot{N}_{\text{per-MBH}}\,{\propto}\, M_\bullet^{B}$ with $B\approx 0.7$ for the mass range $M_\bullet = 10^6 -10^8 \Modot$ and $M_\ast = 10^8 - 10^{11} \Modot$.\par
% We show a relevant application in Appendix \ref{sec:WD} on the rates of the disruption of white dwarfs.
%galaxy distribution behavior by age

%Comparison with Pfister 2020 and other work
Our model predictions in Fig.~\ref{fig:pergalaxy} are then compared with the ones of \cite{Pfister20} which, motivated by the computation of TDE rates in observed galaxy profiles, computed the per-galaxy TDE rates for a mock catalog of galaxies hosting NSCs and with a well-characterized bulge component. Before making a comparison, it is worth noticing that \cite{Pfister20} derives MBH masses using, also in the low-mass regime,  $M_\bullet - M_{\rm *}$ scaling relations observationally derived for massive spheroidal galaxies \citep[][]{Kormendy13,Davis19}. Moreover, although they use {\phaseflow} to compute the TDE rates, they assume a monochromatic star distribution of $m_\ast \,=\,1\Modot$, while this work uses $m_\ast \,=\, 0.38\Modot$ and stellar black holes ($16 \Modot$), a difference that can change by a factor of two the number of stars entering the loss cone, thus available for TDEs \citep{Stone16}.\par
The per-galaxy rates for NSCs (solid red lines) are consistent with the predictions of \cite{Pfister20} for $M_\bullet\,{>}\,10^{6.5}\Modot$ and $M_\ast\,{>}\,10^{10} \Modot$ but have a different trend (they diverge) at lower masses. We suspect that the main difference to the rates towards lower masses is the use of steep S\'ersic profiles $n_{\rm NSC}$ for low mass NSCs, using the scaling relation of \cite{Pechetti20} (for $M_{\rm NSC} \,{<}\,10^{6}\Modot$ the relation gives S\'ersic indices with values above 4). This anti-correlation between NSC mass and  S\'ersic index for lower masses appears to be weaker in the recent results of \cite{Hoyer23}. We instead assume shallower, possibly more realistic, profiles, as described in Sect.~\ref{sec:phaseflow} and no correlation of the steepness of the profile with mass. Moreover, \cite{Pfister20} do not allow for small structures with $M_\bullet\,{>}\,M_{\rm NSC}$, while there is evidence of such objects and they are naturally included in our model (see Sect.~\ref{sec:extracomp} for the impact of this parameter).\par
Regarding the TDE rates from the galaxy component, \cite{Pfister20} also finds an increase both with MBH and galaxy stellar mass until the event-horizon suppression mass. Nevertheless, the average rate they predict is three to four orders of magnitude higher than our predictions for all galaxy stellar mass ranges. This is probably because \cite{Pfister20} assumes on average more centrally concentrated galaxies compared to our work. Also, as mentioned earlier, at the low galaxy stellar mass end, they assign MBH masses using the same steep relation as for a more massive system: this implies that, at fixed MBH mass, their stellar mass can be up to $2\, \mathrm{dex}$ larger than ours.\\
The results just discussed highlight that the demographical analysis of TDE rates requires a realistic cosmological environment with carefully constructed morphological galaxy properties and number distributions as a function of galaxy stellar mass, that  {\lgbh} provides. 

\subsubsection{Redshift evolution of TDE rates}

Having captured the $z\,{=}\,0$ global rates, we now discuss their redshift evolution and how that compares with the evolution of the underlying MBH properties and their environment.\par

\begin{figure*}
\centering
%VERSION 1 SIMPLISTIC, REDUCED
%\includegraphics[width=0.98\textwidth]{figures/NSCRateTables/fiducial/2byz.pdf}\\
\includegraphics[width=0.98\textwidth]{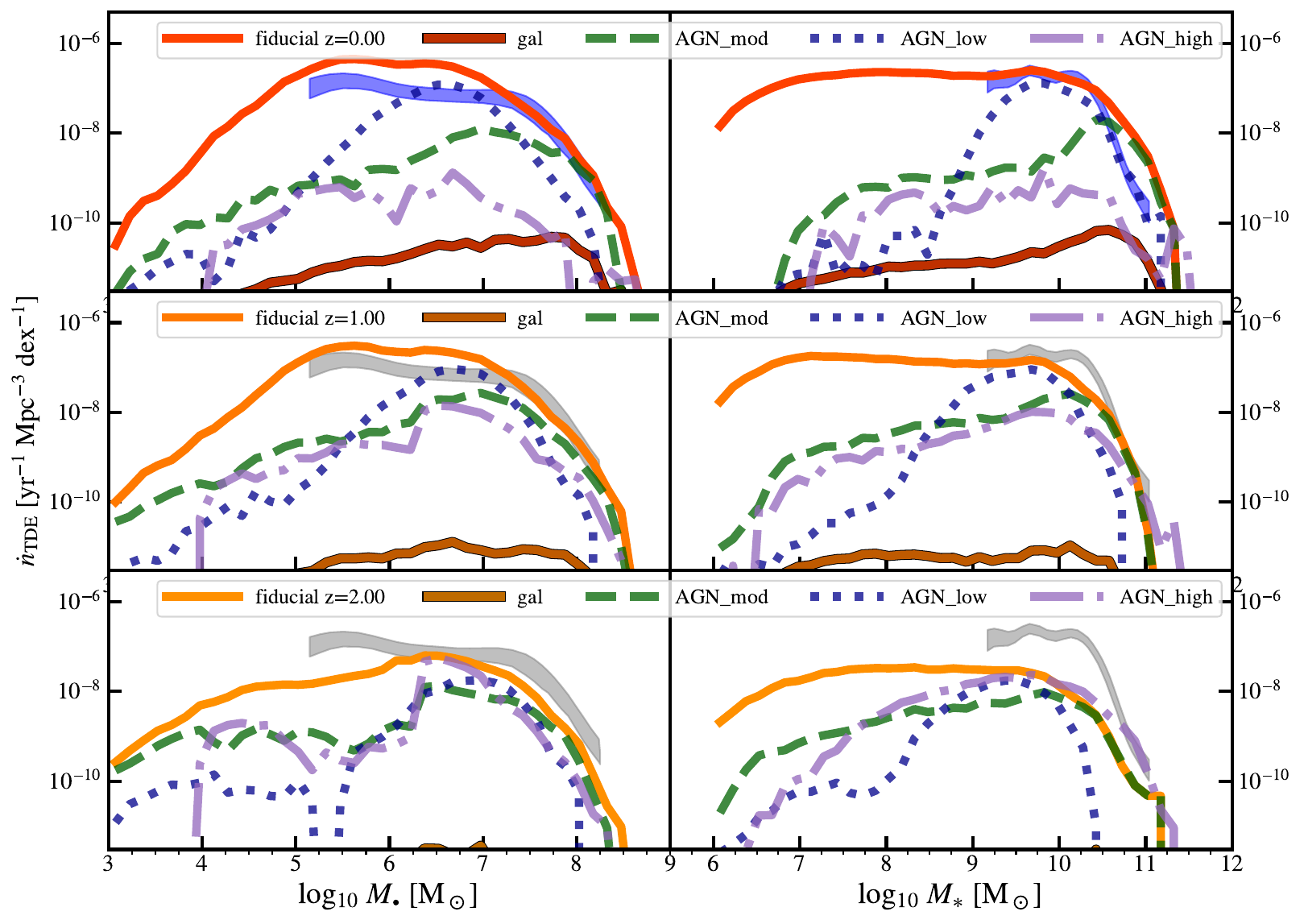}\\
  \caption{Redshift evolution [$z\,{=}\,0$ (\textit{top}), $z\,{=}\,1$ (\textit{middle}), $z\,{=}\,2$  (\textit{bottom})] of the volumetric TDE rates per MBH (\textit{left}) and stellar (\textit{right}) mass originating from {\it all} low-luminosity or inactive MBHs (cut-off at $\log_{10}L_{\rm AGN} \, {\rm [erg/s]}<42$ for the {\it fiducial} model, solid lines) and TDE rates from the galaxy components only (bulges/disks; solid lines tagged as ``gal''). We also display the rates only from AGN-hosts with {\it high} (purple dashed-dotted), {\it moderate} (green dashed), and {\it low} (blue dotted) luminosity ($\log_{10}L_{\rm AGN} \, {\rm [erg/s]}\in [42,45]$, $[40.5,42]$ and $[39,40.5]$ respectively). The $z\,{=}\,0$ constraints \citet{Yao23} are plotted in all panels for reference (with grey when they do not apply). These fractions remain qualitatively the same for these redshifts for most of the parameter variations.} 
\label{fig:volrates_split_types}
\end{figure*}

In Fig.~\ref{fig:volrates_split_types} we display the volumetric TDE rates for redshifts $z\,=\,0,1,2$. Regarding the case of TDE rates as a function of MBH mass, there does not seem to be a significant change in the rates between $z\,{=}\,0$ and $z\,{=}\,1$. This is due to the  very mild evolution evolution predicted by the model for $z\,{<}\,1$. This result is in line with the general assumptions that the volumetric TDE rates do not evolve significantly with redshift (see e.g.\citealp[]{vanVelzen14,vanVelzen18,Yao23} see, however, the approach of \citealp{Kochanek16}). Despite this, there is a slight decrease of half an order of magnitude when comparing resolved TDEs at $z\,{=}\,0$, $1$ and $z\,{=}\,2$, which will be an important feature to study with deep sky surveys such as LSST \citep{LSST22,Bricman20,Bricman23} and UV transient surveys like QUVIK \citep{Zajavcek23}. However, it is worth mentioning that our predictions are model-dependent. We assume that MBHs are always associated to an NSC at all redshifts (see Sect.~\ref{sec:condNSC}). If this was not true at early times, high-redshift TDE rates would be lower. \par

The volumetric rates shown in Fig.~\ref{fig:volrates_split_types} can be divided into contributions from the galaxy component (``gal'') and the NSC. The first, although it steadily increases from $z\,{=}\,2$ to $z\,{=}\,0$ with the increase of bulges in galaxies, still provides a negligible contribution to the volumetric TDE rates at all masses and all redshifts. In practice, the volumetric rates are entirely due to galaxies hosting NSCs (this contribution is not shown in the figure as it essentially coincides with the total ``fiducial'' volumetric TDE rates).
While the per-galaxy rates for NSCs remain the same at high masses towards higher redshifts, there is a slight enhancement (factor less than three) compared to $z\,{=}\,0$ for $M_\bullet \,<\, 10^{6}\Modot$ MBH and $M_\ast\,<\,10^9 \Modot$ host-galaxy stellar masses, because of continuous reset of TDE rates during frequent mergers at cosmic noon. This somewhat counteracts the decrease of the number density of inactive MBHs towards higher redshifts and results in a moderate evolution of the volumetric TDE rates.

\subsection{TDE rates in active MBHs} 

Searches for TDEs in AGN at optical \citep[][ZTF]{Dgany23} and X-ray \citep[][eROSITA]{Homan23} wavelengths have not yielded many representative cases. However, observations of luminous ambiguous nuclear transients \citep{Frederick21,Hinkle22,Oates23}, TDEs and TDE-like flares in ongoing/previous AGN \citep{Li22,Huang23liner,Makrygianni23,Charalampopoulos24}, and changing-look AGN \citep{Ricci22,Zhang22} hint that the two TDE and AGN activity could be related. So far, few theoretical works \citep{Karas07,Chan19,McKernan22,Prasad24} and recent simulations \citep{Ryu23b} have explored the co-existence of TDEs and AGN. Predicting TDEs in AGN hosts is beyond the scope of this paper, as it requires dedicated modeling of the electromagnetic emission of individual events, considering the relative brightness of TDEs and their host. This section overviews TDE occurrence in AGN under the assumptions outlined in Sect.~\ref{sec:agnmodel}, aiding future searches for transient events in AGN.\par

In Fig.~\ref{fig:volrates_split_types} we show the volumetric global rates at $z\,{=}\,0$, $1$ and $2$, for MBHs triggering AGN with {\it low}, {\it moderate} and {\it high} bolometric luminosity bins: $\log_{10}\,(L_{\rm bol} \, [\mathrm{erg/s}]) \in [39.0,40.5]$, $[40.5,42.0]$ and $[42.0,45.0]$. Note that for {\it high} AGN luminosity, the detectable rate of events will depend on the luminosity of individual TDEs. In other words, only TDEs with peak luminosity significantly higher than the AGN luminosity at a given wavelength would allow distinguishing the characteristic time-decay of a TDE light curve, thus singling out the event from AGN stochastic variability or unrelated flaring activity. 
The first thing to notice in Fig.~\ref{fig:volrates_split_types} is that the strong redshift evolution of the AGN luminosity function leaves an imprint on the AGN hosting TDEs. Indeed, the lower the redshift, the greater the importance of inactive MBHs in the volumetric TDE rates. This is because a large fraction of massive black holes consume their high-$z$ gas reservoir and end up in an inactive phase in the low-$z$ Universe.\par
We observe that at high MBH and host-galaxy stellar masses ($M_{\bullet} \,{>}\, 10^8 \Modot$ and $ M_{\ast} \,{>}\, 10^{10.5}\Modot$) and at all redshifts, the volumetric rate of TDEs occurring in luminous AGN (all luminosity bins) is comparable to the one occurring in inactive or low activity AGN ({\it fiducial}) curve and almost redshift-independent at $\mathrm{Gpc}^{-3} \, \mathrm{yr}^{-1} \, \mathrm{dex}^{-1}$. Specifically for redshift $z\,{=}\,2$, the TDE rates of luminous AGN ({\it high}) dominate over the rates of low-activity MBHs at host-galaxy stellar masses $M_\ast\,{>}10^8\Modot$ and $M_\bullet\,\sim\,10^{6.5} \Modot$ MBHs. Therefore monitoring flares in AGN at $z\,{>}\,1$ might be an effective way to identify TDEs, assuming that a significant fraction of TDEs will be brighter than the AGN emission itself. \par
%PLATEAU AND AGN EMISSION
%Finally, the volumetric rates for AGN with {\it moderate} AGN bolometric luminosities could be informative for studies of the late-time plateau of TDE light curves \citep{Brown17,VanVelzen19}, which are detected in 2/3 of the time and have UV luminosities $L_{\rm plat} =10^{41}\,{-}\,10^{43}$erg/s \citep{Mummery23b}. For example, for the $M_\bullet =10^{7.5} M_\odot$ with $L_{\rm plat}=10^{42}$erg/s \citep[see scaling][]{Mummery23b}, assuming that 20\% of the AGN bolometric luminosity goes into the UV band, we predict that 1/10 of the time (independent of redshift) there will be a 1-20\% contribution to the plateau from the AGN accretion flow if it persists after the TDE.
%--------------------------------------------------------------------
\section{Discussion} \label{sec:discussion}

Here we address the implications of the inclusion of TDEs for the modeling of MBHs with {\lgbh}, and explore how the results presented in this work are affected by the parameters and assumptions of our model (Section~\ref{sec:model}).

\subsection{MBH growth via Tidal Disruption Events}\label{sec:growth}

As mentioned in the Introduction, the role of TDEs in the mass content of MBHs has been so far theoretical, with no means to calibrate this mechanism at analytical prescriptions of this MBH growth channel. Since our work already is in broad agreement with the TDE rates at $z\,{=}\,0$, we can draw some first conclusions on the contribution of TDE rates to the growth of the MBH population. Here we limit the discussion to the impact of TDE on the growth of the MBH population by $z\,{=}\,0$. Another important aspect regarding growth is the time resolution; according to \cite{Broggi22} MBHs residing in NSCs will double their mass within a characteristic timescale that depends on the MBH mass (in their set-up, they have a power-law dependence $1.29$) and falls below time resolution for MBHs $M_\bullet <10^5\Modot$. We include the {\it prompt phase} by integrating the rates taking place below the time-resolution $dt_{\rm step}$. A more detailed investigation of the conditions for efficient growth via TDEs and of the sub-categories of galaxy types where and when this growth channel is important will be the subject of follow-up work (Polkas et al.\ in prep).

\begin{figure}
\centering
% \includegraphics[trim={0.0cm 1cm 0cm 1.0cm},clip,width=0.4\textwidth]{figures/additional/averageBHAM_cumul_MBH_z0.0.pdf}
% \includegraphics[trim={0.0cm 0cm 0cm 1.0cm},clip,width=0.4\textwidth]{figures/NSCRateTables/fiducial/averageBHAM_cumul_MBH_z0.0.pdf}
% % \includegraphics[trim={0.0cm 0cm 0cm 0cm},clip,width=0.4\textwidth]{figures/NSCRateTables/fiducial/averageBHAM_cumul_MBH_z0.0_minus_z1.0.pdf}
%   \caption{Average fraction of mass accreted by different MBHs at $z\,{=}\,0$ ($f_{\rm BHAM}$) via stellar captures (purple squares), cold gas accretion (blue right triangles), hot gas accretion (orange left triangles) and accumulated seed mass (green circles), without (top panel) and with (bottom panel) an initial prompt phase (see main text). The margin of $\pm 1$ standard deviation of $\log_{10} f_{\rm BHAM}$ for each type of channel is highlighted with the same-color shaded area.}

\includegraphics[trim={0.0cm 0cm 0cm 1.0cm},clip,width=0.48\textwidth]{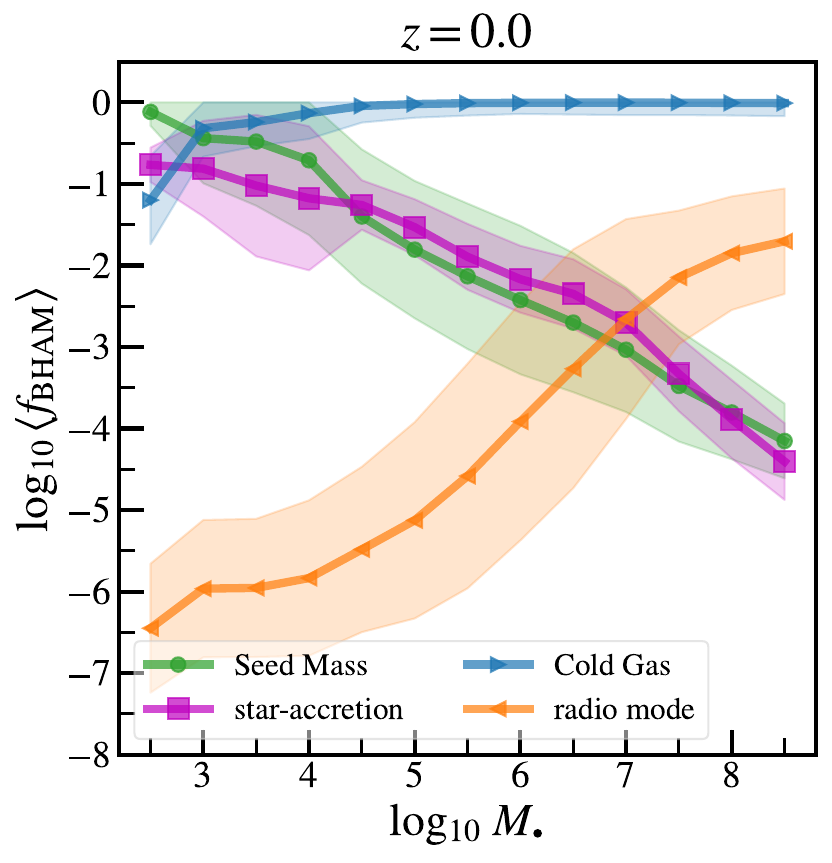}
  \caption{Average fraction of mass accreted by different MBHs at $z\,{=}\,0$ ($f_{\rm BHAM}$) via stellar captures (purple squares), cold gas accretion (blue right triangles), hot gas accretion (orange left triangles) and accumulated seed mass (green circles), without (top panel) and with (bottom panel) an initial prompt phase (see main text). The margin of $\pm 1$ standard deviation of $\log_{10} f_{\rm BHAM}$ for each type of channel is highlighted with the same-color shaded area.}
\label{fig:rel_growth}
\end{figure}

In Fig.~\ref{fig:rel_growth}, we show the fractional cumulative mass content of MBHs, namely $f_{\rm BHAM}$\footnote{Here, the definition of the total black hole accreted mass (BHAM) includes the mass content in seed mass, although strictly speaking, this is not accreted onto the MBH.} at $z\,{=}\,0$ as a function of their mass, for the runs with and without {\it prompt phase}. As demonstrated {\it in these plots}, the mass growth through TDEs is mostly insignificant for the high mass ranges (i.e. $f_{\rm BHAM} \,<\,$1\% for $M_\bullet \,{>}\,10^5 \Modot$), especially when compared to the mass growth induced by cold gas accretion. However, for MBHs that remain close to their seed mass $M_{\bullet}\,=\,10^{2.5}\,{-}\,10^5\Modot$, TDEs offer a competitive channel of growth, composing $f_{\rm BHAM} \,=\,1-10\%$ of their mass. Notably, for the lightest mass bin $M_{\bullet}\,{<}\,10^3 \Modot$, TDE growth is as important as the cold gas accretion, the latter being less significant (if at all existent) in dwarf galaxies \citep[see Fig.~7 and Fig.~10 of][respectively]{IV19,Spinoso23}. Regarding this result, we stress however the theoretical uncertainties regarding the position and residence time of MBH seeds in the centers of galaxies and NSCs that may diverge from the physics included in {\lgbh} regarding more massive MBHs.\par
For the relative massive MBHs at $z\,{=}\,0$ ($M_\bullet\,{>}\,10^6$M$_\odot$), their seeds grow through gas accretion to $M_\bullet=10^{4}\,{-}\,10^{6} \Modot$ by $z\,{=}\,2-3$, allowing for important cumulative TDE growth (after the initial gas growth) until $z\,{=}\,0$. This is shown by the \emph{excess} which brings the star-accretion curve over the seed mass one (for MBHs with $M_\bullet \,{=}\,10^{4.5} \,-\,10^{7.5}\Modot$). Both $M_\ast$ and $M_{\rm NSC}$ will be greater for MBHs of this mass range, resulting in constant high rates with time compared to those of surviving dormant seeds living in isolation. The latter cannot maintain a constant high rate as indicated by the per-MBH rates for \textit{old} NSC systems in Fig.~\ref{fig:pergalaxy}. More massive MBHs ($M_\bullet \,{\sim}\, 10^8 \Modot$) form early in the most massive galaxies ($M_\ast \,{\sim}\, 10^{11} \Modot$) of the simulation which will stop hosting NSCs earlier and grow less their MBH through TDEs.

Our results add realism to estimates from earlier works. Unlike claims using the full loss-cone theory \citep[e.g.][]{Milosavljevic06,Bar-Or17}, the growth via TDEs in our model is not such that all $M_\bullet \,{<}\, 10^4\Modot$ MBHs would reach the supermassive regime as a result of stellar accretion. We attribute this behavior to a) the vast majority of low-mass MBHs being hosted in lower-mass galaxies with less massive NSCs and b) the time-dependent nature of the rates; non-interacting ancient NSCs relax quickly and can even become less important than the bulge/disk rates (see input rates Fig.~\ref{fig:rates_and_rhos}). As mentioned above, however, we will study in a follow-up work the specific properties of the environment and redshifts that lead to more efficient growth via TDEs.

\subsection{Implications for MBH and NSC populations}\label{sec:implMBHNSC}

\begin{figure}
\centering
\includegraphics[width=0.48\textwidth]{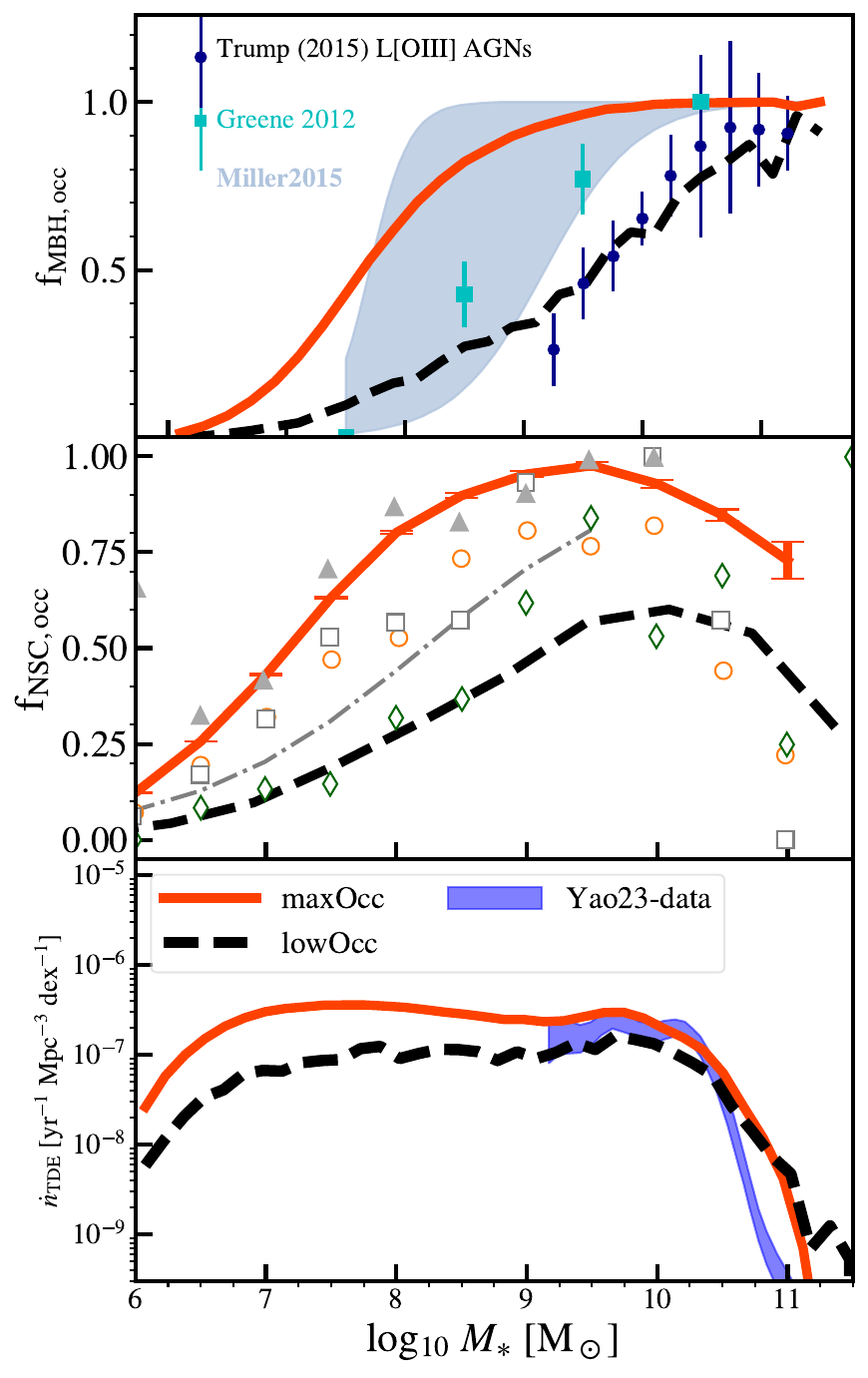}
\caption{{\it Top and middle panels:} occupation fractions of MBH ($M_\bullet\,{>}\,10^{5}\Modot$, \textit{top}) and NSC (all masses, \textit{middle}) that result into the volumetric TDE rates as a function of galaxy-host mass (\textit{bottom}). We show the {\it maxOcc} and {\it lowOcc} models with orange solid and black dashed lines, respectively. In the top panels we display the local MBH occupation fraction derived from a compilation of detections \citet[][cyan squares]{Greene12} and X-ray constraints by \citet[][light-blue region]{Miller12}, as well as an occupation for slightly higher redshifts $0.01\,{<}\,z\,{<}\,0.1$ derived via optical-line AGN \citep[][blue circles]{Trump15}. In the middle panel, the data points and logistic function (the dashed-dotted thin line), both from \citet{Hoyer21}, have the same meaning as in Fig.~\ref{fig:lgbh-nucleation}. The enhanced MBH occupation is still compatible with the observations of the Coma Cluster (grey triangles), even though it yields a greater fraction of nucleated galaxies. We stress that all NSCs are hosting an MBH in our model, hence adding NSCs without MBHs would increase this fraction.{\it Bottom panel:}
Both {\it maxOcc} and {\it lowOcc} model (solid and dashed line) reproduce the TDE-rate distribution constraint of \citet{Yao23} within errors. However, in the lack of evidence of TDEs in dwarf galaxies the {\it lowOcc} model is preferred.}
\label{fig:comptwomodels}
\end{figure}

Complete galaxy samples are starting to constrain the AGN occupation fraction across different galaxy stellar masses, hinting at significantly smaller AGN fractions in dwarf galaxies (i.e. ${<}\,10^{-2}$) than %the one reported 
in massive systems \citep[see e.g.][]{Mezcua18,Mezcua20,Bykov23,Zou23,Siudek23a}. This could be explained if the number of AGN in dwarf galaxies remains small because of the lack of cold gas in the center of these systems \citep{Urquhart22} which cannot sustain significant MBH growth. Therefore, AGN cannot be used as good tracers of low-mass MBHs (nor of the full population of MBHs). Alternatively, the abundance of MBHs in the local universe and beyond can be unveiled by exploiting the population of TDEs in all-sky transient surveys.\par
As shown in Section~\ref{sec:bhlgbh} and Section~\ref{sec:condNSC} our {\it fiducial} model is in fair agreement with both the MBH mass function and the NSC occupation fraction. However, there are uncertainties at the low-mass end of the MBH/host-galaxy stellar mass function and the frequency of MBHs hosted by NSCs is currently unknown. To address these uncertainties, we compare our {\it fiducial} model to a run with the highest number of MBHs permitted by our model, hereafter {\it maxOcc} model. %prev-may They differ in the amplitude of the MBH seeding probability,} $$\rm G_{p}} \,{=} 0.25 \, (fiducial) \;{\rm vs}\; 1.0 \, (maxOcc),$$ 
They differ in the amplitude of the MBH seeding probability by one order of magnitude, namely $${\rm G_{p}} \,{=} 0.1 \, (lowOcc) \;{\rm vs}\; 1.0 \, (maxOcc),$$ resulting in different MBH occupation fractions of $M_\bullet\,{>}\,10^5 \Modot$  at $z\,{=}\,0$, as shown in the top panel of Fig.~\ref{fig:comptwomodels}. The large occupation fraction of MBHs\footnote{For the sake of brevity, we do not show the black hole mass function of the \textit{maxOcc} model. This model tends to overpredict by a factor of three the number of MBHs in the mass range $M_\bullet \,{=}\, 10^{6}\,{-}\,10^{7.5} \Modot$, compared to observational constraints at $z\,{=}\,0$ from \cite{Shankar09}. Besides this, the overall shape of the mass function remains relatively similar concerning the \textit{fiducial} model at high masses.} in galaxies has a direct effect on the abundance of NSC. Specifically, the \textit{maxOcc} and {\it lowOcc} models display respectively NSC occupation (central panel of Fig.~\ref{fig:comptwomodels}) up to a factor 1.6 larger and lower than the one shown in the \textit{fiducial} case (see Fig.~\ref{fig:lgbh-nucleation}).\par
The bottom panel of Fig.~\ref{fig:comptwomodels} shows that the changes in the MBH and NSC occupation result in appreciable differences in the observed TDE rates per galaxy. Although the {\it maxOcc} model also shows good agreement with the results of \cite{Yao23} it lies on the upper limits of the observational constraints.
Quantitatively, both of the models suggest: $p\,{ \lesssim}\, -1$ for $dn_\bullet\,{/}\,d\log_{10}M_\bullet \,{\propto}\, M_\bullet^p$, instead of $p\gtrsim 0$, as found by \citealt{Yao23}. Also, both the \textit{maxOcc} and \textit{fiducial} models support the interpretation of \citet{vanVelzen18} for a flat ($\sim\!100\%$) occupation fraction of MBHs down to $\rm M_{\ast} \,{\sim}\, 10^{10}\Modot$, while \textit{lowOcc} model does not (see middle panel of Fig.~\ref{fig:comptwomodels}). The noticable difference in the dwarf galaxy regime ($M_\ast = 10^7 \,-\, 10^{10}\Modot$) is the main driver in boosting the volumetric TDEs at all galaxy stellar masses. Conversely, the fact that we do not detect TDEs at $M_*<10^9\Modot$ may be primarily interpreted as a lower occupation fraction of MBHs than the one predicted by {\lgbh}, indicative of a less efficient seeding mechanism.\par
Regarding the NSC occupation, while the \textit{fiducial} model (see Fig.~\ref{fig:lgbh-nucleation}) matches the observations of Virgo (and Fornax at lower masses) clusters \citep{Munoz15,Eigenthaler18,Sanchez-Janssen19}, the \textit{maxOcc} resembles the higher NSC occupation in the Coma cluster \citep{denBrok14,Zanatta21} and the \textit{lowOcc} that of the Local Volume. Most of the clusters in our model are less massive than their MBH mass (see Appendix \ref{sec:smftde} for mass distribution of NSCs). In particular, only $\approx 30\%\; {\rm and}\; 60\%$ of MBHs, hosted in galaxies with mass $M_\ast \,{\sim}\, 10^{7}\Modot\; {\rm and}\;10^{9.5}\Modot$ at $z\,{=}\,0$, are also found in an NSC more massive than them, namely $\mathcal{D}=M_{\rm NSC}/M_{\bullet}\,{>}\,1$. For reference, the median mass ratio between NSCs and their associated MBHs with a measured mass is $\mathcal{D}\,\sim \,4$ \citep{Greene20}, although detections are made in the most massive systems hosting MBHs $M_\bullet \,{>}\,10^6 \Modot$. \par
We underline that there is a degeneracy between the occupation of MBHs per galaxy and the occupation of NSCs per MBH. In particular, simultaneous increase and decrease of each number density can result in the same observed normalization of the volumetric TDE rates. Interestingly, denser environments (galaxy clusters) are found to have higher-than-average occupation fractions of both MBHs \citep{Tremmel23} and NSCs \citep{Sanchez-Janssen19,Hoyer21,Zanatta21,Leaman22} at a given galaxy stellar mass, in line with {\it maxOcc} model predictions. Based on our model and the novel TDE rates per-galaxy of our study we speculate that the following conditions should be fulfilled by future models, for them to explain the observed TDE rates: i) a tight connection between the two families of nuclear compact objects (in our model all NSCs host an MBH) and ii) the dominant contributor to TDE rates to be MBHs hosted in $M_\ast \,{\sim}\, 10^{9}\,{-}\,10^{10.5}\Modot$ galaxies in dense environments, where NSC and MBH occupation reach 100\% and the per-galaxy rate is peaking. An explicit study of the environment dependence, involving realistic star clusters and NSC modeling will be needed to further understand the connection between the two classes of objects.

\subsection{Spin Evolution \& Implications for TDE rates} \label{sec:Spin_Implications}

\begin{figure}
\centering
\includegraphics[width=0.48\textwidth]{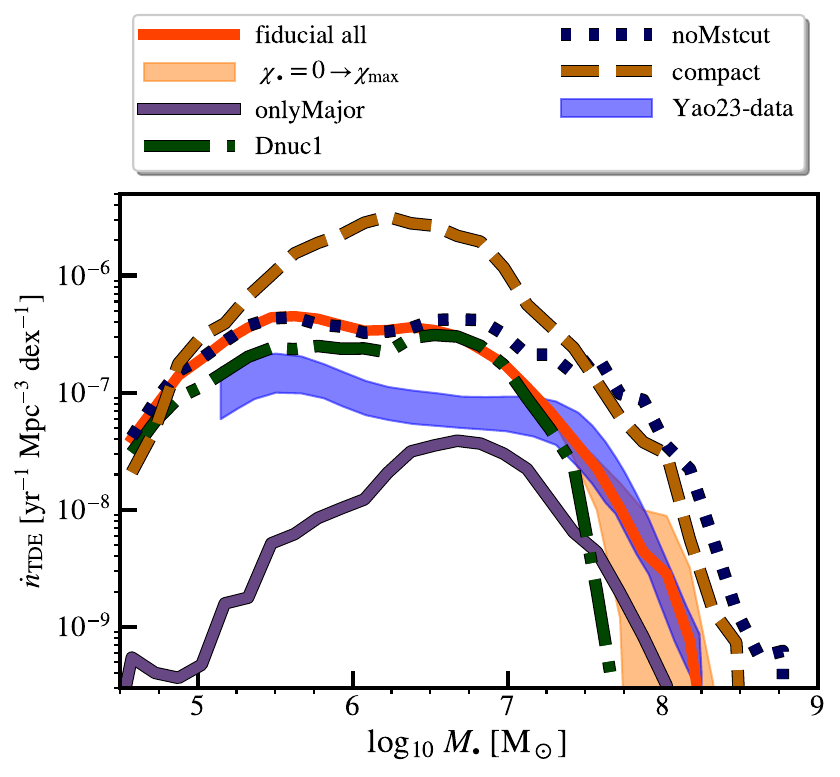}
\includegraphics[trim={0cm 0cm 0cm 0cm},clip,width=0.48\textwidth]{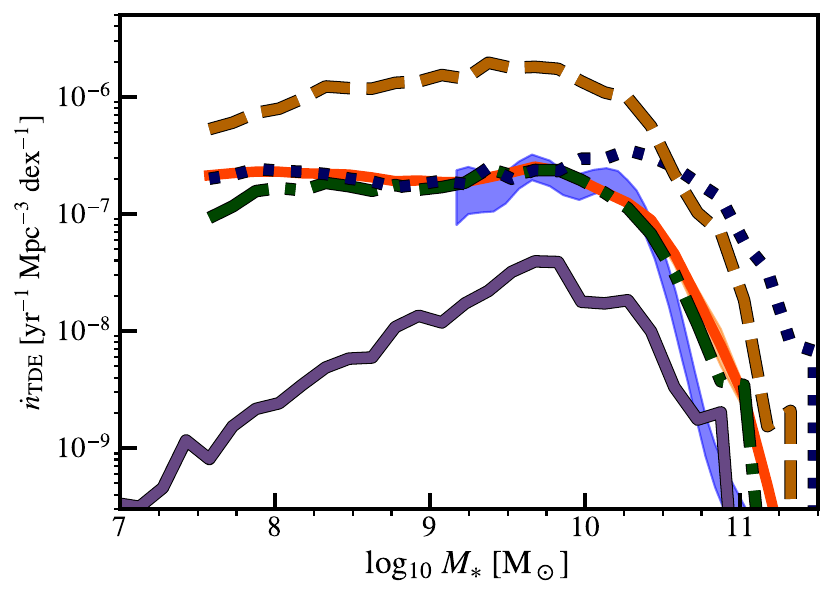}
  \caption{ Volumetric rates at $z\,{=}\,0$ per black hole mass (\textit{top}) and galaxy stellar mass (\textit{bottom}) for different model variations (names of the models in legend as in text), compared with constraints from \citet[][blue shaded area]{Yao23} and the {\it fiducial} model (red solid line). The shaded orange area is constructed by assuming no spin and maximally spinning MBHs in the post-processing treatment of the event-horizon suppression of the {\it fiducial} model (orange solid line).} 
\label{fig:rate_comp}
\end{figure}

MBHs close to the Hills mass ($M_{\bullet} \,{\sim}\, 10^8\Modot$) should exhibit near-to-zero TDEs by $z\,{=}\,0$, and if they do they should be spinning \citep{Kesden12}. Also, because of their rarity, any TDE observed in this mass range will reside statistically at higher redshift \citep[][]{vanVelzen18,Mummery20,Hammerstein23}. Still, with the handful of massive systems observed with TDEs in the ZTF sample \citep{Yao23} we can test large volume statistics of the model's MBH spin distribution.

As shown in Sect.~\ref{sec:baseline}, {\lgbh} predicts a median spin of $\chi_\bullet \,{\approx}\, 0.75$ for MBH with masses $M_\bullet{>}\,10^7\Modot$, with a standard deviation of $\sigma_{\chi_\bullet} \, \approx \,0.2$. Note that when initially applied on the lower-resolution \textit{Millennium-I} simulation \citep{IV20}, the spin model predicts that $M_\bullet {\lesssim}\, 10^8\Modot$ MBHs remain high-spinning ($\chi_\bullet \,{>}\, 0.8$) down to $z\,{\sim}\, 0.1$. Overall, the agreement\footnote{We stress that we have identified the slight underprediction to be due to the lack of massive NSC abundance by the model as discussed in our results (see Sect.~\ref{sec:extracomp}, model tagged as \textit{noMstcut}).} of the shape of the predicted volumetric rates at high MBH masses with observations (Fig.~\ref{fig:globalrates}) suggests that the model makes a reasonable prediction of the spin distribution at $z\,{<}\,0.5$. 

To guide the reader and bracket the effect of the spin model on the event-horizon suppression, Fig.~\ref{fig:rate_comp} displays the volumetric TDEs for the same MBH population of the \textit{fiducial} model but assuming $\chi_\bullet = 0$ and $\chi_\bullet =0.998$ when computing the event horizon suppression (Sect.~\ref{sec:postproc}) as a shaded region around the model line. As shown, a non-spinning MBH population in the mass range $M_\bullet \,{=}\,10^{7.5}\,{-}\, 10^{8.5}\Modot$ (left border of the shaded region) can be excluded while a maximally spinning population in the same range, the model rates lie within the error margins of the constraints. 
Our results are consistent with the recent findings from \cite{Mummery23b} who derived MBH spin $\chi_\bullet\,{=}\,0.3\,{-}\,0.75$ by modeling the late-time TDE emission of 10 MBHs with mass $M_\bullet\,{>}\,10^7 \Modot$.

We are careful with our interpretation since for converting simulated TDE rates to observable ones we have neglected some aspects of the disruption of the stars themselves. For example, we have used a truncated ($m_\ast\,{<}\,1.5\Modot$) Kroupa IMF for the event-horizon suppression calculations. Nevertheless, the scenario of a {\it top-heavy} initial stellar mass functions for the stellar nuclear environment is frequently referenced as an alternative to high-spinning MBHs \citep{Mockler22}. Our results may also be affected by the distribution of the stellar orbit inclinations as recently indicated by \citet{Singh24}. Furthermore, the effects of the De Sitter precession of the streams of the disrupted star on the outcome of a TDE may play an important role\footnote{in the scenario where radiation predominantly comes from stream collisions} \citep[stream self-crossing becomes ineffective][for reference]{Bonnerot22,Jankovic23}. Nevertheless, our model suggests that basic assumptions for the star population and the detectability of individual events (Sect.~\ref{sec:postproc}), coupled with the spin distribution of {\lgbh}, can fairly reproduce the observed event-horizon suppression.

\subsection{The impact of parameters choice}\label{sec:extracomp}

As seen before, the occupation fraction of MBHs in galaxies seems to play an important role in the normalization of the volumetric TDEs, MBH spin distribution regulates the shape at the event horizon suppression. Here, we examine how TDE rates are influenced by other free parameters related to the NSC model (nucleation, regeneration, and compactness) and by the conditions used to reset the TDEs. We highlight that we do not seek to explore the full parameter space but rather pinpoint the most important choices for the model. Also, we stress that our main results for MBH growth do not change qualitatively among runs featuring single-parameter variations.

\paragraph*{\bf More compact NSCs.} 
NSCs are the main stellar environments in which TDEs occur (compared to the galaxy contribution, see both per-galaxy \& volumetric TDE rates in Fig.~\ref{fig:pergalaxy} and Fig.~\ref{fig:volrates_split_types}). When computing the number of TDEs of NSCs inside {\phaseflow} we assumed an effective radius of NSCs according to specific scaling relations (see Sect.~\ref{sec:phaseflow}). However, a large number of uncertainties are still present in these relations. For example, NSCs hosting an MBH are expected to dynamically evolve (evaporate/expand), so the observed scale radius may be greater than the initial radius \citep{Merritt09}. Therefore, we have also explored how the compactness of the NSC affects the number of TDEs and consequently, the volumetric TDE rates (hereafter {\it compactNSC} model). In Fig. \ref{fig:rate_comp}, we show the volumetric rates for a run where NSCs are initialized with a scale radius $a$ reduced to $1/3$ of what is assumed in our {\it fiducial} model (see Sect.~\ref{sec:grid}). This choice is motivated by observations showing NSCs who lie $\sim 0.5  \, \mathrm{dex}$ below the mass-radius relation for NSCs at $z\,{=}\,0$ \citep[see e.g.][]{Pechetti20}. We observe that the model overpredicts the rates for MBH masses $M_\bullet \,{=}\,10^6\,{-}\,10^{7.5} \Modot$ by more than an order of magnitude. The overestimation disfavors NSCs that are significantly more compact than the observed relations used in our {\it fiducial} model. We stress that with the current implementation of NSCs, the model lacks the most massive NSCs of mass $M_{\rm NSC}\,{=}\, 10^{8}\,{-}\,10^{9}\Modot$ which are naturally more compact and will contribute at TDE rates of $M_\bullet\,{>}\,10^{7.5}\,\rm M_{\odot}$ MBH.\par

\paragraph*{\bf Variation of $\mathcal{D}_{\rm nuc}$.} 
This is a key parameter in our model since it controls the frequency of NSC regeneration. Considering lower values than our {\it fiducial} case, observations presented in \citep{Neumayer2020} suggest a value of $\mathcal{D} \,{\equiv}\, M_{\rm NSC}/M_{\bullet}$ as low as $0.01$. However, it remains unclear if these configurations are just transients. Despite that, we test a value of $0.01$, and no significant changes were found. For the sake of brevity and clarity, we do not display the results of this run. Instead, we boost $\mathcal{D}_{\rm nuc}$ towards higher values, from $0.1$ to $1$ (hereafter {\it Dnuc1} model) and show our results in Fig.~\ref{fig:rate_comp}. For this case, the rates are heavily suppressed at high MBH mass ($M_\bullet \,{>}\,10^{7.5}\Modot$) since the NSCs with values $\mathcal{D}_{\rm nuc} \,{=}\, 0.1\,{-}\,1$ are not nucleated in this run. This is a piece of evidence that small and possibly unresolved NSCs ($\mathcal{D}<1$) are the main contributors to TDE rates in the event-horizon suppression mass range. However, only the replacement of this parameter with a physical NSC treatment (including the destruction of NSCs) can offer hindsight on why this is the case.\par

\paragraph*{\bf Variation of cut-off galaxy stellar mass for nucleation.} 
Recently, the careful observational sampling of massive galaxies presented in \cite{Ashok23} finds a possible higher nucleation fraction in the high-mass end than the one reported in the observational constraints used for this work \citep[i.e.][]{Hoyer21}. In our phenomenological model of NSCs, we impose a maximum galaxy stellar mass $M_{\rm *, NSC-cutoff} \,{=}\,10^{9.75}\Modot$ (Sect. \ref{sec:condNSC}), above which either the conditions are not adequate to trigger the NSC formation or, equivalently, the destruction of NSC becomes important. The specific value of this cut-off is extremely important for $M_\ast\,{>}\,10^{11}\Modot$  galaxies since higher values of the cut-off will both allow for a higher number density of NSCs, but also more massive NSCs (since the scaling relation extends to high-mass galaxies). On the other hand, a sufficiently low $M_{\rm *, NSC-cutoff}$ can turn off the nucleation around $M_\bullet\,{>}\,10^8\Modot$ MBHs, converting the galaxy component to the only source of TDEs in this mass range.\par

To explore how the volumetric rates are affected by the nucleation mass cut-off, in Fig.~\ref{fig:rate_comp} we present the results of a run where no cut-off is applied, i.e all galaxies at all masses can host an NSC as long as  $\mathcal{D}_{\rm nuc}\,{>}\,0.1$ (hereafter, {\it noMstcut} model). As compared to the {\it fiducial} model, this parameter variation shows that the inclusion of more massive clusters at more massive galaxies won't affect the low-mass regime of the volumetric rate distribution and increase the rates at the event horizon suppression with error margins. These ``more massive'' NSCs can be similarly introduced by increasing the scatter on the $M_\ast \,{-}\,M_{\rm NSC}$ scaling relation when assigning NSC masses in the model. We anticipate the inclusion of a realistic scenario for NSC evolution and destruction will increase the TDE rates in massive galaxies.\par

\paragraph*{\bf Variation of TDE rates reset frequency.}
To assess the importance of the conditions used to reset the TDE rates (see Sect.~\ref{sec:reset}) we consider that only major events are capable of resetting the clock of TDEs ($\Delta t_{\rm TD,i}$). In particular, major events refer here to (i) major galaxy mergers (satellite/central baryonic mass ratio ${>}$10\%) and (ii) disk instability events that displace more than ${>}$10\% of the stellar disk mass to the bulge. Hereafter, we refer to this model as {\it OnlyMajor} and we present its volumetric TDE rates in Fig.~\ref{fig:rate_comp}. Overall, the predictions of the model are ${\sim}\,1\, \rm dex$ below the current observational constraints regardless of the MBH and galaxy stellar masses. Yet, the suppression of rates induced by the {\it OnlyMajor} model for galaxies with $M_\ast \,{\sim}\, 10^{11}\Modot$ leads to a better agreement with observations than for the {\it fiducial} model. This points out that minor events occurring in massive galaxies should take place away from the galactic nucleus, hindering their possibility of resetting the TDE rates. While this might be true for massive systems, it is unlikely that for small galaxies catastrophic events such as major mergers or important disk instabilities are the unique processes able to significantly modify the stellar environment around nuclear MBHs and thus reset the relaxation process.

The underprediction of the volumetric TDE rates for small MBHs and galaxies shown in {\it OnlyMajor} model highlights that NSC structures (that dominate the overall TDE rates, see Sect.~\ref{sec:VolumetricTDEsResults}) should be dynamically influenced by their host galaxy interactions quite often. For instance, the accretion of non-major satellites and weak disk instabilities should result in a relatively frequent refilling of the loss cone of their hosted MBH. An important caveat to this argument is that catastrophic events are the type of interactions in which the basic assumptions of spherical symmetry and isotropy of stellar profiles (assumed in this work) are most frequently violated and a rate enhancement may occur (see Sect.~\ref{sec:lim_and_prosp}). \par

%The non-asymmetries or overdense galactic nuclei that occur after this kind of event could lead to a very plausible enhancement in the TDE rates concerning the ones computed from the standard symmetric and isotropic conditions (see further discussion in Sect.~\ref{sec: enhancement}).

%%%%%%%%%%%%%%%%%%%%%%%%%%%%%%%%%%%%%%%%%%%%%%%%%%%%%%%%%%%%%%%%%%%%%%%%%%%%%%%%%%%
\section{Limitations and Prospects}\label{sec:lim_and_prosp}
Here we discuss the main limitations which cannot be addressed with our methods. In addition, we discuss further details on the nature of TDE events which could potentially be captured in the framework of our model. We make some suggestions for improvements regarding both the use of {\phaseflow} and {\lgbh}, as well as motivate further research projects as an extension to this work.

\subsection{Caveats regarding input TDE rates}

As pointed out in \citet{Merritt13}, the 1D Fokker Planck approach (see Sect.~\ref{sec:phaseflow}) may be inaccurate over the relaxation timescale. A more general treatment that accounts for the evolution of energy and angular momentum may be reliable on longer timescales, over which the angular momentum profile alters the overall evolution of the stellar profile \citep[see e.g.\ ][]{Broggi22}. Moreover, there are effects that cannot be taken into account in {\phaseflow}, at least in its current form. Noticeable ones are: resonant relaxation of Keplerian orbits \citep{Rauch96}, large scattering \citep[instead of diffusing into the loss-cone][]{Weissbein17} or loss-cone shielding \citep{Teboul22}; all of which have been shown to have an impact on TDE rates for specific systems. On top of this, relativistic effects may affect the outcome of the interactions and thus the prediction of the rate \citep[e.g.][]{Stone19}. Addressing all these caveats would require a different solution to the problem than the one provided with two-body energy relaxation as performed by {\phaseflow}. \par
In addition to the limitations due to the negligence of anisotropies, the current method is limited to working with spherical systems (while many systems are observed rotating and therefore flattened). Although \cite{Merritt13} argues that non-axisymmetry affects rate estimation within a factor of two, we stress that rates for axisymmetric profiles have been noted to deviate from the ones produced with {\phaseflow} \citep[see e.g.][]{Vasiliev13}.\cite{Kim18} found evidence of elongated bulges being correlated with nuclear starbursts\footnote{This relates to the open questions of the effects of stellar bars on central nuclear activity.}, which may be a possible way to allow an increase in rates for a specific type of bulges. The bursty behavior of high TDE rates in certain galaxy types has been explained through radial anisotropies \citep{Stone18}, occurring through the formation of eccentric disks \citep{Madigan18,Rantala24b}, as opposed to overdensity in the nucleus (as assumed in this work). Nevertheless, it is worth noticing that observations have been favoring the latter \citep{French20b}. All these are subjects that have mainly been explored by expensive N-body simulations with tailored initial conditions and go far beyond the general/average viewpoint we have drawn for the global TDE rates.\par
The stellar initial mass function is also an important caveat to expand on. All systems (\textit{young} and \textit{old}) are assigned TDE rates from an evolved population of stars, represented by the average of the Kroupa mass function, and heavy black holes. The particle mass and the fraction of total mass of this heavy component ($16 \Modot$ at $7\%$ of the cluster's mass respectively) drive the magnitude of mass segregation in energy \citep[a process that has its own timescale and may damp the rates, see][]{Bortolas22}. Since the initial phase is affected by mass segregation and low-mass systems clusters during the initial phase have higher per-galaxy TDE rates (see Fig.~\ref{fig:pergalaxy}), the choice of these parameters should be investigated further in the future when studying the growth and the rates at these systems.\par
Regarding the complexity of the initial mass function, the theoretical calculation of rates only demands the first two moments of mass \citep[see][]{Stone16}, therefore we expect that a more complete mass distribution will not impact severely the total number of events (but only a factor of a few). Moreover, in our model, the NSC profiles are motivated by the assembly of NSCs through the accumulation of globular clusters \citep[][]{Antonini12}, which will have themselves evolved stellar populations \citep[see recent advancement in simulations of the formation of NSCs by][]{Gray24}. However, for clusters that form for the first time in a galaxy at high redshift, the assumed actual initial mass function could be lacking stellar black holes that need some time to form as remnants of the first massive stars, but also it could be more top-heavy than is currently assumed \citep{Chon21,Sneppen22,Cameron23}. Going beyond the bi-chromatic distribution into a realistic initial mass function is limited for now from the computational feasibility of a huge parameter space for the input tables \footnote{The computational cost scales with the number of stellar particle mass bins times the number of different initial mass functions we would like to test.}.
%\footnote{As we have noted in Sect.~\ref{sec: discussion} and is shown in the NSC mass function in Appendix \ref{sec:smftde}, the most massive clusters which via observations have been detected with in-situ star-formation, are not reproduced within our model.}

\subsection{Caveats regarding the galaxy environment}

Regarding the galaxy evolution model used in this work, one of the main aspects affecting the TDE demographics is the galaxy sizes predicted by {\lgbh} \citep[initially introduced by][]{Guo11}. Despite being traced self-consistently the expected sizes do not capture some particular galaxy classes discovered in the last decade (e.g.\ ultra-compact dwarfs). The inclusion of more refined environmental effects included in other {\lgfamily} versions \citep[e.g.][that adds ram-pressure stripping]{Ayromlou21}, may be necessary to capture the nature of these types of objects.\par

Regarding galaxy profiles, there are also some improvements in the modeling of stellar dynamics that can be included in our semi-analytical approach. Specifically, the S\'ersic indices assigned to bulges by {\lgbh} are motivated by $z\,{=}\, 0$ observations. However, this does not take into account the evolutionary history of galaxies. For instance, systems in a post-starburst and early quiescent phase \citep{Setton22}, as well as ``red nuggets'' \citep[e.g.][]{Saracco10,Barro13,Ito23,Baggen23,Lohmann23,Pandya24}, have been shown to display a larger Sérsic index than ordinary galaxies. Finally, the bulge velocity dispersion is not included self-consistently in {\lgbh}. This hinders the possibility of exploring the effect of this quantity on the TDE rates. For instance, observations of ``$\sigma$-drop'' bulges \citep{Comer08} suggest that some bulges will have a low enough velocity dispersion to significantly shorten relaxation timescales and thus change the expected TDEs \citep{Cen20}.\par

Regarding the reset mechanism (Sect.~\ref{sec:reset}), internal processes such as phase-mixing, gravitational heating from perturbers \citep[e.g. giant molecular clouds,][]{Merritt13}, and bar interactions with the nuclear region will certainly affect the event rates. These effects cannot be captured by our model since they are not quantified in {\lgbh} but rather would require modeling through expensive, dedicated hydrodynamical simulations that also include TDEs.\par 
Finally, regarding the NSC phenomenological scheme itself, we do not explore nor the redshift-dependence of the scaling relations neither the conditions under which a galaxy will host an NSC. For example, NSCs can both lose mass over time due to stellar evolution, two-body relaxation, tidal shocks and MBH interactions and gain mass through \textit{in-situ} star formation or cluster mergers. This makes uncertain the evolution of the NSC-galaxy stellar mass scaling relation in Eq.~\ref{eq:mnscscale}. If we assume that, at higher redshifts and in dwarf galaxies, NSCs were more massive and more compact than the observed values at $z \sim 0$, the TDE rates will increase accordingly and could result in a disagreement with the observational constraints. Conversely, suppose we assume that most high-mass NSCs in the present-day Universe accumulated their mass over time via \textit{in-situ} star formation. In that case, they might have been lower-mass at higher redshift compared to today, which would result in lower high-redshift TDE rates (causing a stronger redshift-dependence in Fig.~\ref{fig:volrates_split_types}) and could contribute to the overprediction of TDE rates in massive galaxies (right panel of Fig.~\ref{fig:globalrates}).

\subsection{Rate enhancement}
The highest per-galaxy rate predicted by our \textit{fiducial} model corresponds to ${\sim}\,10^{-4}$yr$^{-1}$, compatible with the earliest results of \cite{Syer99}.
Despite that, a sample of galaxies like ULIRGs or E+A(k+a) post-starbursts seem to feature rates of $10^{-3}$ to $10^{-1}$ per year, values that are not reproduced in any individual galaxy of {\lgbh}, and rarely occur in the input grid of TDE rates (see Sect.~\ref{sec:grid}). The large rate seen in post-starburst galaxies could be the result of a TDE enhancement driven by particular conditions. For instance, the recent theoretical work of \cite{French17} supports the amplification of TDEs in the core of post-starburst galaxies, driven primarily by the compactness of these systems and not by the initial stellar mass function variations \citep{Bortolas22}. Another interpretation is that as soon as the AGN phase is over, the loss cone region around the MBH can be filled, thus boosting the expected TDE rates by up to two orders of magnitude compared to standard nuclei \citep{Wang23c}. Moreover, a significant fraction of MBHs in {\lgbh} are found in binaries \citep[in MS by z=0 galaxies with $M_* = 10^{10}-10^{11} \Modot$ host MBH binaries with primary's mass $M_\bullet >10^7 \Modot$, see Fig.~8 in][]{IV23} at which the rates of the secondary MBH can be enhanced to be similar to those of the most massive one \citep[see][and discussion further below]{Li17,Ryu22,Mockler23}. There are therefore three plausible explanations for the scarce number of cases with TDEs rates per MBH as high as $\rm {>}\,10^{-4}\, yr^{-1}$ in our model\footnote{These explanations rely on the assumption that high TDEs rates per MBH (${>}\,10^{-3}\, yr^{-1}$) are not due to observationally unresolved NSCs (discussed later in this section).}: i) the stellar profile of dwarf and intermediate-mass galaxies is not correctly captured at all times, ii) at least one enhancement mechanism due to additional physics is determining the global rates and should be included, or iii) the theoretical uncertainty regarding the initial NSC profiles and the accuracy of the initial relaxation phase hinders the possibility of probing the initial prompt phase of violent relaxation for these objects. To examine the last explanation, a cautious approach for the initial ``bursty'' phase when resetting the stellar environment around the MBH (see Sect.~\ref{sec:phaseflow} \&~\ref{sec:reset}) may resolve this problem. This will be the subject of exploration in future work. \par
If the rates in the aforementioned over-represented classes of galaxies are indeed dominated by NSC rates, a more careful treatment of NSCs at different galaxy types could yield exceptionally high rates. Denser, cuspier, and more massive clusters are all conditions that have a non-negligible impact on the evolution of time-dependent TDE rates \citep{Broggi22}. We have already noted that the introduced phenomenological model does not produce the most massive NSCs of mass $M_{\rm NSC}\,{=}\, 10^{8}\,{-}\,10^{9}\Modot$ (see the mass function in Appendix~\ref{sec:smftde}). In addition to that, the formation channel via accretion of clusters (assumed in this work) appears to operate in dwarf galaxies ($M_{\rm NSC}\,{<}\,10^{8}\,{-}\,10^{9}\Modot$, \citealt{Johnston20,Fahrion21,Fahrion22}); yet central star formation may dominate cluster formation in larger galaxies \citep{Aharon15,vanDonkelaar23}. Consequently, the spheroidal profile used for this work could not be applied in such cases (see also \citealt{Kacharov18} and \citealt{Pinna21} for differences with host galaxy morphology). However, for the inclusion of cuspier profiles, mass segregation in angular momentum---which cannot be captured by {\phaseflow}---will prevent the model from computing correctly the TDE rates and a more sophisticated approach may be needed.

\subsection{Other classes of star-MBH interactions:}

In this work, we have focused our results on the rates of full TDEs and their contribution to MBH growth. However, there are many possible directions for future work exploring physically different types of interactions between stars and MBHs.

\subsubsection{Partial Tidal Disruption Events}
  Partial tidal disruption events (partial TDEs) might be equally important for the growth of black holes as full TDEs. For example, \cite{Zhong22} find that partial TDEs are super-Eddington 2.5 times more frequently than full TDEs, while the repetitive nature of these events can spoon-feed the MBH with increased efficiency \cite{MacLeod12,Ryu20c}. Partial TDE rates are as high as those of full disruptions both theoretically \citep{Krolik20} and from recent observations \citep{Somalwar23}. Recently, \cite[][]{Bortolas23} by tracing the fate of the remnants of such events and the inferred rates using {\phaseflow} showed that partial TDEs can be more frequent than full TDEs up to a few tens of times for nucleated galaxies. However, the radiative signatures of partial TDEs, and how they compare with those of full TDEs, need to be addressed before discussing further their statistics \citep[e.g. a TDE can be falsely attributed as a rising AGN event, see e.g.][]{Chen21}. If indeed partial outnumber full TDEs, the volumetric TDE rates may be even lower, possibly indicating a model-observation tension towards the low-mass regime ($M_\bullet < 10^6 \Modot$). One of the dimmest events of \cite[][]{Yao23}, AT2020vdq with an assigned mass of $M_\bullet \,{\sim}\, 10^{5.5} \, \Modot$,  has recently been flagged as a partial disruption \citep{Somalwar23a}, a class of events that is not included in our predictions. If these and other events at $M_\bullet \, < \,10^6 \Modot$ are indeed partial disruptions that would mean a tension between our model and observations (in Fig.~\ref{fig:globalrates}). 
  We continue this discussion quantitatively in Appendix ~\ref{sec:typeTDE} for the interested readers, where we introduce some relevant definitions.\par
In general, different orbit distributions will result in different frequencies of full TDEs, partial TDEs, and direct captures \citep{Park20,Cufari22}, but also different rates for certain subcategories of full TDEs, e.g. relativistic TDEs \citep{Ryu23,Seoane23} and TDEs with eccentric debris disks \citep{Zanazzi20,Wevers22}. Tapping into the statistics of the orbits can serve as an additional time-dependent output in the grid-like exploration of the parameter space which then can be provided as {\lgbh} input, addressing the aforementioned issues. We plan to improve this model aspect in future works.

\subsubsection{Rates from binary and wandering MBHs}
The rich physics shaping the evolution of two (or more) MBHs toward coalescence following galaxy mergers, may significantly affect TDE rates. In the pairing phase, MBHs move within the potential of the bulge (not lying anymore in the dense center) and produce different rates of TDEs than central MBHs, with possible enhancement of rates during passages of stellar overdensities \citep[][]{Merritt13}. After the formation of a binary (on parsec scales), the eccentric Kozai-Lidov mechanism can boost the rates of the secondary MBH to be similar to the most massive one \citep{Chen11,Mockler23}. Also at smaller separations (a multiple of the sum of the two MBHs tidal radii), the rates are theoretically expected to be boosted \citep{Li17,Ryu22}. An important enhancement would allow for an indirect inference of binaries through TDEs \citep{Mockler23}, something to be investigated in follow-up work dedicated to rates from binary MBHs.\par

Finally, the inclusion of wandering MBHs, assuming that a significant amount of them can maintain a massive star cluster after ejection, may contribute to the overall TDE signal and add to the off-nucleus signal from MBHs in globular clusters \citep[expected to be lower than the nuclear MBHs][]{Ramirez-Ruiz09}, especially when considering the smallest MBHs. The inclusion of such rates could lead to a possible explanation of the non-detection of a host galaxy \citep[e.g. the long-duration TDE ``scary barbie''][]{Subrayan23} or other off-nuclear transients \citep[e.g. the Fast Blue Optical Transient ``Finch''][]{Chrimes24}.

\subsubsection{Electromagnetically-dark events}

Extreme Mass Ratio Inspirals (EMRIs) and Direct Plunges \citep[as defined in][]{Broggi22} of stellar remnants have their own evolution within an NSC. Given the relatively large mass accreted for each EMRI and plunge event, MBH growth via dark accretion of compact objects might even be more important than accretion through TDEs. However, while the distribution of stars in NSCs can be constrained by observations, there is no direct way to probe the distribution of non-luminous objects\footnote{This is the main reason these events were not included in our analysis because they do not fall into the same category as the optically detected TDEs. Yet, our model is already capable of investigating these phenomena, provided that it we track a population of stellar black holes within the NSC.}: this adds large theoretical uncertainties in calculating the cosmological rates. Also, regimes where general relativity effects are dominant over two-body relaxation should be taken into consideration.\par 
Also of great interest, binary stars have a large tidal radius \citep[10-100 times that of a single star]{Merritt13}, while a binary system may be disrupted consecutively, either by one or both MBHs \citep{Wu18}. For realistic EMRI and plunge rates, triple interactions between stellar remnant binaries and the MBH must be considered \citep{Bonetti19,Bonetti20}. Our method is currently not ideal for the inclusion of binary stars and remnants. 
%Since TDE rates may depart from our predictions, this class of events may alter our results for high binarity in NSCs and galaxies. % Notably, the newest version of {\lgfamily} includes the stellar evolution code of binary stars \citep{Yates23}, which may allow for further development in the future. 
We aim to investigate the aforementioned type of interactions, along with their contribution to the gravitational wave background to be probed by Laser Interferometry Space Antenna (LISA), in a future dedicated study.

%%%%%%-----------------------------------------------------------------------

\section{Conclusions}\label{sec:conclusions}

In this work, we have included for the very first time tidal disruption events (TDEs) from massive black holes in a semi-analytic model of galaxy formation and evolution. To this end, we have used the {\lgfamily} model \citep{Henriques15} in the version presented in \cite{IV20,IV22} and \cite{Spinoso23}, which includes many physical processes that drive the formation and evolution of massive black holes (MBHs). For this work, we also included a phenomenological model for nuclear star clusters (NSCs) that uses the observed NSC - galaxy stellar mass relations and reproduces the observed NSC occupation fraction.  Time-dependent TDE rates for each galaxy are then obtained by solving the 1D Fokker-Planck equation with {\phaseflow} for a variety of stellar environments surrounding the MBHs. Finally, we have made simple post-processing assumptions for transforming theoretical to observable rates. The key findings of the model are summarized in the following:

\begin{itemize}
\item[$\bullet$] Our model predicts volumetric TDE rates that are in agreement with the observed ones. Rates from disruptions of stars belonging to the stellar disk or bulges alone can not explain observations. Instead, we predict that the majority of TDEs should take place in NSCs, independently of galaxy or black hole mass. To explain the rates within the local cosmological volume ($z\,{<}\,0.5$), a steep black hole mass function (${\propto}\, M_\bullet^{-1}$ at $M_\bullet \,{\sim}\, 10^5\,{-}\,10^{6.5} \Modot$) and a relatively high occupation fraction of black holes also in the dwarf regime are needed.

\item[$\bullet$] The TDE rates per-MBH do \textit{not} follow a single power law as obtained in other works from scaling relations \citep[e.g. $\dot{N}_{\text{per-MBH}}\, {\propto} \, M_\bullet^{B}$ with $-0.4 \,{<}\, B\,{<}\, 0$, see][]{Wang04,Stone16,Pfister20}. Instead, our model produces a positive power-law dependence, with $B \,{\approx}\, 0.7$ for the TDE rates from the galaxy component that peak at $M_\bullet \,{=}\, 10^{8}\Modot$, before the event-horizon suppression completely takes over. For the rates from the NSC component (which dominate the total rates), we predict a power law index of $B\,{=}\,0$ and $B\,{=}\,1$ for \textit{young} and \textit{old} NSCs, respectively, turning over at $M_\bullet \,{\sim}\, 10^{7} \, \Modot$. 

\item[$\bullet$] The spin distribution, a result of the MBH spin model introduced in {\lgbh} by \citet{IV19}, naturally explains the event horizon suppression observed in the optical TDE samples. The model points out that MBHs of mass $M_\bullet \,{=}\, 10^7\,{-}\,10^{8.5} \, \Modot$ have median spin $\bar{\chi}_\bullet \,\approx\,0.75$ with a standard deviation of $\sigma_{\chi_\bullet}\,\approx\,0.2$.

\item[$\bullet$] A great fraction of TDEs ($\sim$100\% for the most massive hosts $M_\ast\,{>}\,10^{11} \Modot$) takes place around MBHs which are also experiencing some level of gas accretion. We thus predict TDEs to be detectable on top of low-luminosity AGN (bolometric luminosity $3 \times 10^{40}$-$10^{42} \, \mathrm{erg/s}$), although AGN are generally excluded from TDE searches. Also, we predict TDE-like flares to have a volumetric rate of $0.3-1\, \mathrm{Gpc}^{-3} \, \mathrm{yr}^{-1} \, \mathrm{dex}^{-1}$ for Seyfert galaxies with AGN luminosity $L_{\rm bol} \,{=}\, 10^{42}$-$10^{45} \, \mathrm{erg/s}$ in the broad mass range $M_\ast \,{=}\, 10^{8}-10^{10.5} \, \Modot$, with the rate increasing by a factor of a few tens at redshifts $z\,{\ge}\,1$.
\end{itemize}

To conclude,  the model  presented here, a combination of {\lgbh} and {\phaseflow}, can overall reproduce the latest observed TDE rates, while respecting the constraints on the galaxy and black hole mass functions.
Our results highlight the importance of using a realistic cosmological framework to obtain a comprehensive view of TDE demographics. Furthermore, our results underline the importance of including time-dependent TDE rates \cite[as demonstrated in][]{Broggi22}, which allow us to account for the relaxation timescale in the statistics, unlike studies that use instantaneous TDE rates.\par
Finally, while we do not find a significant contribution of TDEs to the average MBH mass growth,  TDE growth might be important for specific classes of MBHs and seeds, which will be investigated in a follow-up work (Polkas et al.\ in prep). We also plan to address the caveats and limitations of our model in future works to improve its ability to draw predictions for TDEs and the associated growth of MBHs, in particular in view of the upcoming flow of data from LSST.

\begin{acknowledgements}
This work used, as a foundation, the 2015 public version of the Munich model of galaxy formation and evolution: {\lgfamily}. The source code and a full description of the model are available at \href{https://lgalaxiespublicrelease.github.io/}{https://lgalaxiespublicrelease.github.io/}. \\
It also uses the publically available 1D-Fokker Planck solver {\phaseflow} available at \href{http://eugvas.net/software/phaseflow/}{http://eugvas.net/software/phaseflow/}. \\
We thank Rosa Valiante for kindly allowing us to use her catalogues of \texttt{GQd} for the PopIII sub-grid physics. \\
M.P. and S.B. acknowledge support from the Spanish Ministerio de Ciencia e Innovación through project PID2021-124243NB-C21.\\
A.S.,  D.I.V. and E.B. acknowledge the financial support provided under the European Union’s H2020 ERC Consolidator Grant ``Binary Massive Black Hole Astrophysics'' (B Massive, Grant Agreement: 818691).\\
E.B. acknowledges support from the European Union's Horizon Europe programme under the Marie Skłodowska-Curie grant agreement No 101105915 (TESIFA), from the European
Consortium for Astroparticle Theory in the form of an Exchange Travel Grant, and the European Union’s Horizon 2020 Programme under the AHEAD2020 project
(grant agreement 871158).\\
N.H.\ is a fellow of the International Max Planck Research School for Astronomy and Cosmic Physics at the University of Heidelberg (IMPRS-HD). N.H.\ received financial support from the European Union's HORIZON-MSCA-2021-SE-01 Research and Innovation programme under the Marie Sklodowska-Curie grant agreement number 101086388 - Project acronym: LACEGAL.\\ D.S. acknowledges the support of the National Key R\&D Program of China (grant no.\ 2018YFA0404503), the National Science Foundation of China (grant no.\ 12073014), the science research grants from the China Manned Space Project with No. CMS-CSST2021-A05, and Tsinghua University Initiative Scientific Research Program (No. 20223080023).
\end{acknowledgements}
%%%%%%-----------------------------------------------------------------------

\bibliographystyle{aa.bst} % style aa.bst
\bibliography{biblio.bib} % your references Yourfile.bib

\begin{appendix} %First appendix

\section{TDE hosts mass functions}\label{sec:smftde}

\begin{figure}
\centering
\includegraphics[width=0.48\textwidth]{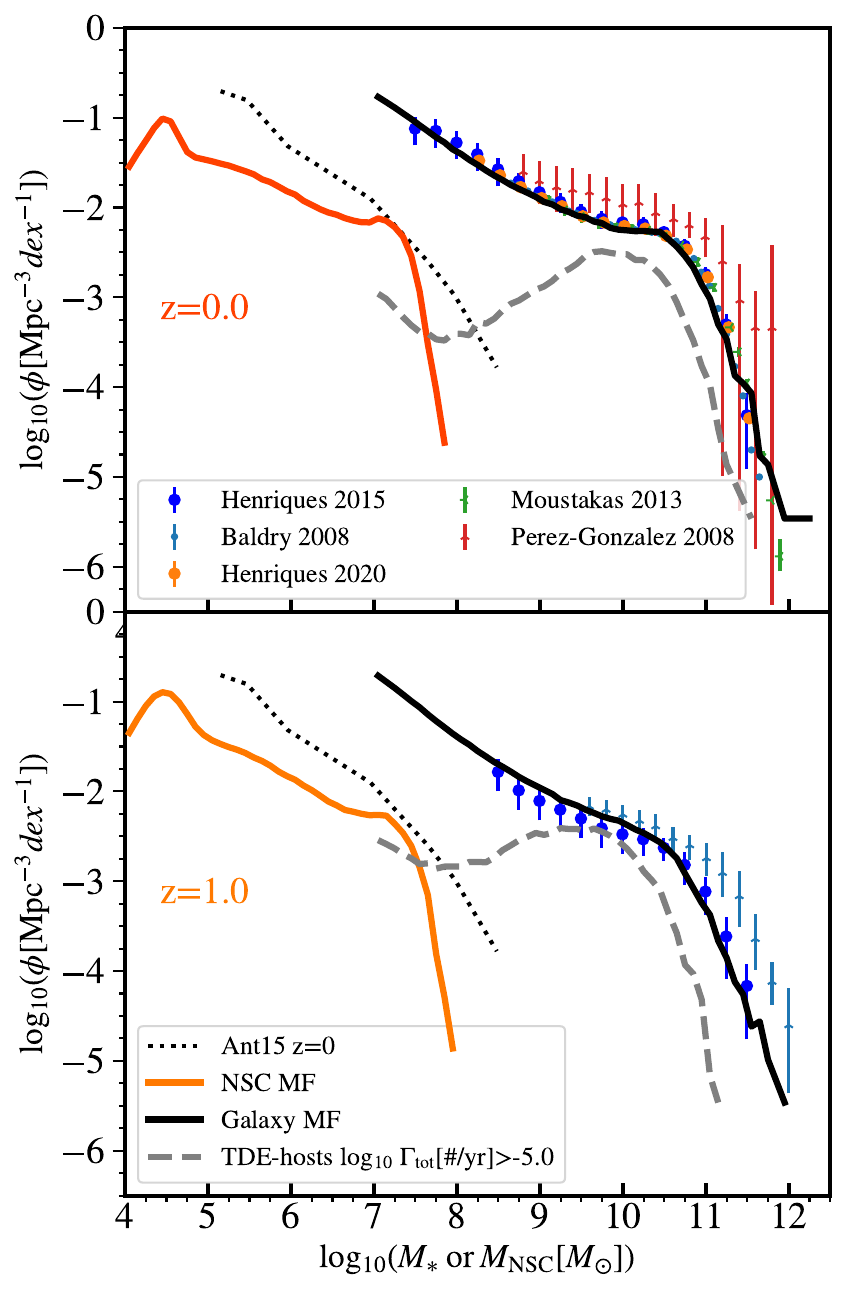}
\caption{Stellar Mass Functions of all galaxies (solid black) and TDE-host galaxies with the highest TDE rates ($\Gamma_{\rm tot}{>}\,10^{-5}$, grey dashed) for the {\it fiducial} model at $z\,{=}\,0$ and $z\,{=}\,1$. The first is compared with compiled mass constraints from \citet{Henriques15} and \citet[]{Henriques20}, \cite{Baldry08}, \cite{Perez-Gonzalez08}, and \cite{Moustakas13} (inset legend). With a thin colored line at lower masses, the NSC Mass Function of the {\it fiducial} model is plotted at the same redshifts. For comparison, we display the model NSC mass function from \cite{Antonini15} at $z\,{=}\,0$ (Ant15 z=0).}
\label{fig:smftde}
\end{figure}

Our model can explain the constraints of \citet{Yao23} on the galaxy stellar mass distribution of the volumetric rates. We discuss briefly the reasons for the relatively good agreement and the issues to be addressed regarding the TDE host mass range in future works.\par
As observed in Fig.~\ref{fig:smftde} (grey dashed), we reproduce the flat galaxy stellar mass function expected for TDE host galaxies as inferred from observations \citep{Law-Smith17}. Galaxy stellar mass functions are all in good agreement with one another and with observations at $z\,{=}\,0$, $1$, as expected by the success of the {\lgfamily} models.\par

In the same plot, we present the NSC mass function at $z\,{=}\,0$ (dotted-dashed) which is lower compared to the theoretical predictions of \citet{Antonini15} by an order of magnitude at lower masses, obviously having a different slope. The number density provided by this function should be a theoretical lower limit to reproduce the TDE rates, yet NSCs without MBHs should also be considered. The model predicts that NSCs do not evolve significantly with redshift from $z\,{=}\,1$, as they are tightly related to the nuclear MBHs (that also do not evolve).\par

At $z\,{=}\,0$ the peak at hosts with high rates within $M_\ast\, =\, 10^{9.5}$\,-\,$10^{10.5} \, \Modot$ is directly related to hosting massive NSC $M_{\rm NSC}\,=\, 10^7$-$10^8 \, \Modot$ and recently seeded (high per-galaxy rates, left panel in Fig.~\ref{fig:smftde} for NSC distribution) while the evolving tail ($M_\ast\,{<}\, 10^9 \, \Modot$ from $z\,{=}\,1$ to $z\,{=}\,0$ is from the contribution of intermediate-mass MBHs and smaller NSC (right NSC distribution in Fig.~\ref{fig:smftde}), and systems with TDE rates decaying with time. The model predicts mild redshift evolution, a falsifiable argument with the observations of TDEs up to $z\,{=}\,1$ from the forthcoming LSST samples.

\section{Useful definitions for Tidal Disruption Event rate interpretation} \label{sec:typeTDE}

An interaction between a massive black hole and a star reaches its proximity zone of less than one tidal radius $r_t$ (Eq.~\ref{eq:r_t} in the main text), at fixed masses $M_\bullet$ and $m_\ast$ respectively, will result in a variety of outcomes, with different energetics and observable signatures depending on many parameters. 
How close to the horizon the star is destined to make a passage is of particular importance \citep{Carter82,Stone13,Dai15}. 
It is useful to quantify this with the penetration parameter that is defined as 
$$\beta =  r_{t}/r_p,$$
where $r_p$ the pericenter of the orbit and $$r_t = 6.9\times 10^{12}{\rm cm}\; \eta_{\rm TD}^{2/3} (M_{\bullet} / 10^6 M_\odot)^{1/3}m_\ast^{-1/3}r_\ast $$ the tidal radius depends on the mass/radius of the star disrupted $m_\ast$ and $r_\ast$ (units of $1 \Modot$ and $1$R$_\odot$). The coefficient $\eta_{\rm TD}$ depends on the polytropic index of the star and is of the order unity \citep{Merritt13}. Penetration factor values range from $\beta_{\rm min}\, =\, 0.5$-$0.6$ as derived from hydrodynamical simulations \citep{Guillochon13} and at which point stars are merely disrupted, to $\beta_{\rm max}$ at which point stars enter whole into the MBH's horizon. Analytical calculation on horizon suppression for a Schwarzchild MBH yields \citep{Kesden12,Mummery23c} \begin{equation}
\beta_{\rm max} = 11.8 M_{\bullet,6}^{-2/3}m_\ast^{-1/3}r_\ast.
\label{eq:bmax}
\end{equation}
Also, below a value $\beta_c$ the star is not fully disrupted. Therefore, full TDEs in the interval $\beta_{\rm c} \,{<}\,\beta \,{<}\,\beta_{\rm max}$ while direct captures occur for $\beta \,{>}\,\beta_{\max}(M_{\bullet})$. The rest of the events are found in the interval $\beta_{\min} \,{<}\,\beta_c$ and are considered partial TDEs. The latter are identified by the re-flaring of their source and less frequently from their differentiation in the fading power law ($-9/4$ instead of $-5/3$). \citet{Ryu20} has shown that the physical tidal radius at which the star is fully disrupted has both stellar and MBH mass dependence. For disrupted stars of mass 0.3 and 1 M$_\odot$, we adopt
\begin{equation}
\beta_{\rm c} = \beta_{\rm c,0}\frac{1}{1+C_{0}(M_{\bullet}/10^6 M_\odot)^{2/3}}.
\label{eq:betac}
\end{equation}
where $\beta_{\rm c,0} =2.0$,$C_{0}=0.166$ \& $\beta_{\rm c,0} =2.65$, $C_{0}=0.204$ respectively. We set the simple form $\beta_{\rm c,0}(m_\ast) = 1.6(m_\ast+0.45)$ and fix $C_0=0.2$ for our simple calculations here (although there is not a linear transition between stars with different polytropic indices). For reference, \citet{Guillochon13} and \citet{Mainetti17} values from simulations for convective \& radiation-dominated stars around a $M_\bullet  = 10^6\Modot$ MBH translate into $\beta_{\rm c,0}=1.08$ \& $2.2$ when fixing $C_0=0.2$ respectively. 
Note that there is a continuous transition from TDEs to direct captures as  $\beta_{\rm c}(m_\ast)\rightarrow \beta_{\max}(M_{\bullet},m_\ast,\chi_{\bullet},t),$ with the Hills mass being rather a range of values for the distribution of star masses. \emph{This transition is computed solely in the analysis here}; in the main text TDEs are considered as $\beta \,{>}\,\beta_{\rm c}=1$ (all stars within the tidal radius are disrupted) with $\beta_{\rm max} \rightarrow \infty$ for $M_{\bullet} \,{<}\, 10^{8}\Modot$ (direct captures are not considered separately) and $\beta_{\max} \,{<}\, \beta_{\rm c}$ for higher MBH masses (all events are considered direct captures).

The number of orbits per penetration factor bin $dN(\beta)/d\beta$ takes the analytical form of a power-law for the full loss-cone regime (the system just started to relax). By assuming uniformity of the orbits over $r_p$ gives a power-law dependence of $\,{\propto}\, \beta^{-2}$ \citep{Stone20}. When the relativistic gravity is taken into account, this power law drops to $\beta^{-10/3}$  \citep{Coughlin22} for maximally-spinning MBHs (Kerr metric). Also, we consider the scenario where orbits are distributed uniformly over surface $2\pi r_p dr_p$ and the probability function scales as $\,{\propto}\, \beta^{-3}$ \citep{Bricman20}. At the empty loss-cone regime (the system has relaxed) there is a weak (logarithmic) dependence of rates on $\beta$, so essentially we can assume $\,{\propto}\, \beta^0$ for this case. For simplicity, we assume that the distribution of orbits destined to do partial TDEs follows from one of the full TDEs as inferred for the disruption of normal stars from \citet{Zhong22}.\par

%Of course, there are additional uncertainties regarding the properties of the star in calculating this limit (e.g. stage, magnetic field, metallicity) that are not addressed here. 

We can now write the fraction of full TDEs as the integral of a power-law probability $\beta^{-S_{\beta}-1}$ in the value intervals mentioned above:
%(related to the size of the loss cone):
\begin{equation}
f_{\rm FTDE}(m_\ast,M_\bullet) = \frac{ \beta_{\rm c}(m_\ast,M_\bullet)^{-S_\beta}- \beta_{\max}(m_\ast,M_\bullet)^{-S_\beta}}{\beta_{\min}^{-S_\beta}- \beta_{\max}(m_\ast,M_\bullet)^{-S_\beta}} %\left(\equiv \frac{J_{\rm c}-J_{min}}{\Delta J_{lc}}\right)
\label{eq:betaprob}
\end{equation}
where the generic power-law $S_\beta$ is in the range $S_\beta \in [1,2]$ for our calculations, for a non-empty loss cone (for all types of events). For the empty loss-cone regime, $S_\beta = -1$ and events at all $\beta$ should be equally rare. 

Now, we can estimate a general correction for our calculated rates of full disruption stars to the rate of all disruptions by using Eq.\ref{eq:betaprob}, for the regime where $\beta_{\rm min} \,{\gg}\, \beta_c \,{\gg}\, \beta_{\max}$ get that full/total TDE rate converges as $f_{\rm FTDE} \rightarrow (\beta_{\rm min}\,{/}\,\beta_{\rm c})^{S_\beta}$. For $\beta_{\rm min}=0.6$ and $M_\bullet \,{\lesssim}\, 10^6 \Modot$ yields almost invariably\footnote{This refers to both convection-dominated and radiation-dominated stars since the fraction of full TDE rates is already is over-estimated by a factor of $(\beta_c)^{S_\beta}$ when assuming the TDE rates of 0.38 M$_\odot$ stars hold for more massive stars {\phaseflow}. Also, the weak dependence of $\beta_c$ on MBH mass in the regime is neglected.} a correction factor of $$1/f_{\rm FTDE} \in [2,4]$$ for $S_\beta \in [1,2]$. This correction could be applied only to \textit{young} systems relaxing only for the last $<100 \, \mathrm{Myr}$ that have not emptied their loss-cone. However, the physics of circularization could be different for TDEs varying in $\beta$, especially for the unknown regime of lower MBH masses \citep[][$M_\bullet  = 10^{2}\,{-}\,10^{4}\Modot$]{Kiroglu22} and the rates may be modified accordingly \citep{Wong22}. These simulations\footnote{Both studies use general relativistic hydrodynamics and initial conditions from the stellar code MESA.} show a critical value greater than the one adopted from studies at $M_\bullet \,=\,10^6 \Modot$, namely $\beta_{c}\,{\sim}\, 10\,{/}\,3 \,{>} 47\,{/}\,19.1$, making the fraction of full disruptions even smaller and this correction greater. Furthermore, this correction factor can be a few tens ($>1 \, \mathrm{dex}$), as demonstrated by the self-consistent work on partial disruption events by \cite{Bortolas23}. By visual inspection, in the optical sample of \cite{Yao23} there are more than a few light curves that show a re-flaring activity, instead of a monotonic drop of luminosity. Caution should be drawn when comparing model event rates with samples of TDEs. In the case that indeed events with $\beta<1$ are included in the observational sample, the predictions of our model would be $>1$dex apart from observations and a more thorough discussion should be done on the optimistic assumptions we made when translating {\phaseflow} rates to observations (see Sect.~\ref{sec:postproc})

\end{appendix}

\end{document}